\documentclass[preprint,showpacs,preprintnumbers,amsmath,amssymb,nofootinbib,showkeys]{revtex4-1}

\usepackage{graphicx}
\usepackage{dcolumn}
\usepackage{amssymb}
\usepackage{mathrsfs}
\usepackage{amsmath}
\usepackage{epsfig}
\usepackage{hhline}
\usepackage[utf8]{inputenc}
\usepackage{color}
\usepackage{multirow}
\usepackage[T1]{fontenc}% for symbol ''>''
\usepackage{verbatim}
\usepackage{graphicx}
\usepackage{slashed}
\begin{document}

\title{Polarization of top and chargino from stop decay in natural SUSY}

\author{Lei Wu}
\author{Hang Zhou}

\thanks{zhouhang@njnu.edu.cn}
\affiliation{School of Physics Science and Technology, Nanjing Normal University, Nanjing, 210023, China}

\begin{abstract}

Top squark (stop) plays a crucial role in the naturalness of supersymmetry (SUSY) models, which has been extensively searched for in the LHC experiments. Due to the parity-violating couplings, top quark and chargino thus produced from stop decay are expected to be polarized. We investigate the polarization of top quark and chargino, from which we find that the mass differences between neutralino and stop/chargino can affect the degree of polarization of top quark/chargino in the stop decay. We then examine the top polarization effects on kinematics of stop decay in natural SUSY (NSUSY) at the LHC and the results show that several observables have the discriminating power for different polarized top quark at the LHC, which may be useful for testing the nature of stop couplings in NSUSY.

\end{abstract}
%by optimizing the signal selection criteria, and can hence provide information about stop mixing

\maketitle

%\tableofcontents

%\newpage

\section{Introduction}
After the discovery of the Higgs boson~\cite{Aad:2012tfa,Chatrchyan:2012xdj}, the next primary goal for the LHC experiments will be searching for new physics beyond the Standard Model (BSM). One of the major theoretical motivations in this direction is the naturalness principle, which implies that new physics should appear around TeV scale. Among various models of this new physics, supersymmetry is the most appealing candidate. It provides an elegant solution to the naturalness problem by introducing supersymmetric particles (sparticles) to cancel the quadratic divergence in Higgs mass. Among a plethora of sparticles, the stop plays a key role in the cancelling. Therefore, searching for stop is an urgent task for testing naturalness in SUSY.

So far, the ATLAS and CMS collaborations have performed extensive searches for stop, including the gluino-mediated stop production and the direct stop pair production. The analysis strategies strongly depend on assumptions of spectrum and nature of sparticles. For example, when $ m_{\tilde{t}_{1}} \gg  m_t + m_{\tilde{\chi}^0_1}$, the top quark from stop decay $\tilde{t}_{1} \to t \tilde{\chi}^0_1$ will be energetic. The $t\bar{t}$ background can be suppressed by using endpoint observables~\cite{Lester:1999tx, Bai:2012gs, Cao:2012rz} and boosted technique~\cite{Plehn:2012pr,Plehn:2011tf,Plehn:2010st}. While in the compressed region with $ m_{\tilde{t}_1} \approx m_t + m_{\tilde{\chi}^0_1}$, the kinematics of stop pair events are very like top pair background. The ISR/FSR jet is usually utilized to reduce such a background~\cite{Hagiwara:2013tva}. Besides, stop can also be singly produced via the electroweak interaction, such as the associated production of the stop and chargino, which will lead to distinctive mono-top/bottom signature at the LHC~\cite{Goncalves:2014axa,Hikasa:2015lma,Goncalves:2016nil,Duan:2016vpp,Cacciapaglia:2018rqf}. On the other hand, if the flavor-conserving two body decays of the stop are kinematically forbidden, the stops produced near the threshold will live long enough to form bound states, i.e. stoponium, which subsequently decay through annihilation into the SM final states~\cite{Drees:1993yr,Martin:2008sv,Barger:2011jt,Duan:2017zar}. Besides traditional cut-flow based analysis methods, machine learning (ML) technique may provide a more powerful way to enhance the LHC sensitivity of stop~\cite{Abdughani:2018wrw}.

It should be noted that the top quark produced from stop decay is polarized due to the parity-violating interactions of stop. The polarization effects can be measured from the kinematics of the decay products~\cite{Krohn:2011tw,Godbole:2015bda,AguilarSaavedra:2012xe,Gopalakrishna:2010xm,Berger:2012an,Cao:2010nw,Bernreuther:2015yna,Kobakhidze:2014gqa,Cao:2011hr}, which can be utilized to reveal the features of stop and other non-SM couplings of the third generation quarks~\cite{V.:2016wba,Abrahantes:2014qta,Belanger:2012tm,Wang:2015ola,Low:2013aza,Shelton:2008nq,Barducci:2012sk}. For example, the top polarization was used to probe the mixing in the stop sector at future lepton colliders~\cite{Kitano:2002ss}. In Ref.~\cite{Perelstein:2008zt}, the effective stop mixing angle was proposed to determine the degree of top polarization. In Ref.~\cite{Bhattacherjee:2012ir}, the polarization observables for boosted top quark from stop decay were studied. It was also shown that the longitudinal polarization of the top quark from stop decays $\tilde{t}_1 \to t \tilde{\chi}^0_{1}$ depends on the masses and mixing in both the stop and the neutralino sectors. Like the top quark, the chargino produced from stop decay process $\tilde{t}_1 \to b \tilde{\chi}^+_1$ is also polarized so that the chargino polarization can be obtained from measuring the angular distribution of its decay products. Since the decay of chargino depends on both the chargino and neutralino mixing matrices, the chargino channel can be used to probe a different set of mixing parameters from those probed by the top channel. In this paper, we will investigate the polarization of top quark and chargino from stop decay, and examine the impact of top polarization on kinematics of stop decay in the natural SUSY at parton and detector levels. This paper is organized as follows. In Section II, we calculate the spin-analyzing power of polarized top quark and chargino. In Section III, we investigate several kinematic observables for different polarized top quark in the process $pp \to \tilde{t}_1 (\to t \tilde{\chi}^0_1) \tilde{t}^*_1 (\to \bar{t}\tilde{\chi}^0_1) $ in natural SUSY at the LHC. Finally, we draw our conclusions in Section IV.

\section{Polarized top quark and chargino from stop decay}

In the MSSM, the mass-squared matrix of top-squark in the basis of gauge eigenstates ($\tilde{t}_{L},\tilde{t}_{R}$) is given by
\begin{align}
M^{2}_{\tilde{t}}=\begin{pmatrix}m^{2}_{\tilde{t}_{L}} && m_{t}X^{\dag}_{t}\\ m_{t}X_{t} && m^{2}_{\tilde{t}_{R}}\end{pmatrix},
\end{align}
where
\begin{eqnarray}
m^{2}_{\tilde{t}_{L}}&=&m^{2}_{\tilde{Q}_{3L}}+m^{2}_{t}+m^{2}_{Z}\left(\frac{1}{2}-\frac{2}{3}\sin^{2}\theta_{W}\right)\cos2\beta,  \label{mtl} \nonumber \\
m^{2}_{\tilde{t}_{R}}&=&m^{2}_{\tilde{U}_{3R}}+m^{2}_{t}+\frac{2}{3}m^{2}_{Z}\sin^{2}\theta_{W}\cos2\beta,\label{mtr} \nonumber \\
X_{t}&=&A_{t}-\frac{\mu}{\tan\beta}.\label{xt}
\end{eqnarray}
Here $m_{\tilde{Q}_{3L}}$ and $m_{\tilde{U}_{3R}}$ are the soft-breaking mass parameters for the third generation squarks, respectively. $A_{t}$ is the top-squark trilinear soft-breaking parameter. We neglect the mixing among generations in our study. The mass eigenstates of stops $\tilde{t}_{1}$ and $\tilde{t}_{2}$ are related to the gauge eigenstates by a unitary matrix,
\begin{align}
\begin{pmatrix} \tilde{t}_{1}\\ \tilde{t}_{2} \end{pmatrix} =\begin{pmatrix} \cos\theta_{\tilde{t}}&& \sin\theta_{\tilde{t}}\\ -\sin\theta_{\tilde{t}} && \cos\theta_{\tilde{t}} \end{pmatrix} \begin{pmatrix} \tilde{t}_{L}\\ \tilde{t}_{R} \end{pmatrix},
\end{align}
in which the stop mixing angle $\theta_{\tilde{t}}$ is given by,
\begin{eqnarray}
\sin2\theta_{\tilde{t}}=\frac{2m_{t}X_{t}}{m^{2}_{\tilde{t}_{1}}-m^{2}_{\tilde{t}_{2}}}.\label{stt}
%\cos2\theta_{\tilde{t}}={}&\frac{m^{2}_{\tilde{Q}_{3L}}-m^{2}_{\tilde{U}_{3R}}+m^{2}_{Z}\left(\frac{1}{2}-\frac{4}{3}\sin^{2}\theta_{W}\right)\cos2\beta}{m^{2}_{\tilde{t}_{1}}-m^{2}_{\tilde{t}_{2}}}\label{ctt}.
\end{eqnarray}
%Then, we can obtain stop masses at tree level,
%\begin{eqnarray}
%m^{2}_{\tilde{t}_{1}}&=&m_{t}X_{t}\sin2\theta_{\tilde{t}}+\frac{1}{2}\left(1+\cos2\theta_{\tilde{t}}\right)m^{2}_{\tilde{Q}_{3L}}+\frac{1}{2}\left(1-\cos2\theta_{\tilde{t}}\right)m^{2}_{\tilde{U}_{3R}},\\
%m^{2}_{\tilde{t}_{2}}&=&-m_{t}X_{t}\sin2\theta_{\tilde{t}}+\frac{1}{2}\left(1-\cos2\theta_{\tilde{t}}\right)m^{2}_{\tilde{Q}_{3L}}+\frac{1}{2}\left(1+\cos2\theta_{\tilde{t}}\right)m^{2}_{\tilde{U}_{3R}}.
%\end{eqnarray}

The relevant couplings between stop and electroweakinos in the mass eigenstates are given by,
\begin{eqnarray}
\mathcal{L}_{\tilde{t}_{1}\bar{b}\tilde{\chi}_{m}^{+}}&=&\bar{b}\left(f^{\tilde{\chi}^{+}}_{L}P_{L}+f^{\tilde{\chi}^{+}}_{R}P_{R}\right)\tilde{\chi}^{+C}_{m}\tilde{t}_{1}+\textrm{H.c.}         \label{bt1chi}\\
\mathcal{L}_{\tilde{t}_{1}\bar{t}\tilde{\chi}_{i}^{0}}&=&\bar{t}\left(f^{\tilde{\chi}^{0}}_{L}P_{L}+f^{\tilde{\chi}^{0}}_{R}P_{R}\right)\tilde{\chi}^{0}_{i}\tilde{t}_{1}+\textrm{H.c.}         \label{tt1chi}
\end{eqnarray}
with $P_{L,R}=\frac{1}{2}\left(1\mp\gamma^{5}\right)$ being the projection operators and the coefficients are
\begin{eqnarray}
f^{\tilde{\chi}^{+}}_{L}&=&y_{b}U^{*}_{m2}\cos\theta_{\tilde{t}}, \label{fcharginoL} \nonumber\\
f^{\tilde{\chi}^{+}}_{R}&=&-g_{2}V_{m1}\cos\theta_{\tilde{t}}+y_{t}V_{m2}\sin\theta_{\tilde{t}}, \label{fcharginoR}\nonumber\\
f^{\tilde{\chi}^{0}}_{L}&=&-\left[\frac{g_{2}}{\sqrt{2}}N_{i2}+\frac{g_{1}}{3\sqrt{2}}N_{i1}\right]\cos\theta_{\tilde{t}}-y_{t}N_{i4}\sin\theta_{\tilde{t}}, \label{fchiL}\nonumber\\
f^{\tilde{\chi}^{0}}_{R}&=&\frac{2\sqrt{2}}{3}g_{1}N^{*}_{i1}\sin\theta_{\tilde{t}}-y_{t}N^{*}_{i4}\cos\theta_{\tilde{t}}.\label{fchiR}
\end{eqnarray}
Here $y_{b}=\sqrt{2}m_{b}/(v\cos\beta)$ is the bottom quark Yukawa coupling and $y_{t}=\sqrt{2}m_{t}/(v\sin\beta)$ is the top quark Yukawa coupling. The neutralino and chargino mixing matrices are $N_{ij}$, $V_{mn}$ and $U_{mn}$~\cite{Gunion:1984yn}. By using effective mixing angles, we can rewrite the Eqs.~(\ref{bt1chi}) and (\ref{tt1chi}) as
\begin{eqnarray}
\mathcal{L}_{\tilde{t}_{1}\bar{b}\tilde{\chi}_{m}^{+}}=g^{(\chi)}_{\textrm{eff}}\bar{b}(\sin\theta^{(\chi)}_{\textrm{eff}}P_{L}+\cos\theta^{(\chi)}_{\textrm{eff}}P_{R})\tilde{\chi}^{+C}_{m}\tilde{t}_{1},
\end{eqnarray}
with
\begin{eqnarray}
\tan\theta^{(\chi)}_{\textrm{eff}}=\frac{y_{b}U^{*}_{m2}\sin\theta_{\tilde{t}}}{-g_2 V_{m1}\cos\theta_{\tilde{t}}+y_{t}V_{m2}\sin\theta_{\tilde{t}}}, \nonumber
\end{eqnarray}
and
\begin{eqnarray}
\mathcal{L}_{\tilde{t}_{1}\bar{t}\tilde{\chi}_{i}^{0}}=g^{(t)}_{\textrm{eff}}\bar{t}\left(\sin\theta^{(t)}_{\textrm{eff}}P_{L}+\cos\theta^{(t)}_{\textrm{eff}}P_{R}\right)\tilde{\chi}^{0}_{i}\tilde{t}_{1},
\end{eqnarray}
with
\begin{eqnarray}
\tan\theta^{(t)}_{\textrm{eff}}=\frac{y_{t}N_{i4}\cos\theta_{\tilde{t}}-\frac{2\sqrt{2}}{3}g_1N_{i1}\sin\theta_{\tilde{t}}}{\sqrt{2}\left(\frac{g_2}{2}N_{i2}+\frac{g_1}{6}N_{i1}\right)\cos\theta_{\tilde{t}}+y_{t}N_{i4}\sin\theta_{\tilde{t}}}.\nonumber
\end{eqnarray}
Similarly, the coupling of $W\tilde{\chi}^0_i\tilde{\chi}^+_m$ can also be given by,
\begin{eqnarray}
\mathcal{L}_{W\tilde{\chi}^0_i\tilde{\chi}^+_m}=g_{\textrm{eff}}^{(W)}W^{+}_{\mu}\bar{\tilde{\chi}}^{0}_{i}\gamma^{\mu}\left(\sin\theta^{(W)}_{\textrm{eff}}P_{L}+\cos\theta^{(W)}_{\textrm{eff}}P_{R}\right)\tilde{\chi}^{+}_{m},
\end{eqnarray}
where
\begin{eqnarray}
\tan\theta^{(W)}_{\textrm{eff}}=\frac{-N_{i4}V^{*}_{m2}+\sqrt{2}N_{i2}V^{*}_{m1}}{N^{*}_{i3}U_{m2}+\sqrt{2}N^{*}_{i2}U_{m1}}. \nonumber
\end{eqnarray}

\begin{figure}[t]
\begin{minipage}{0.48\linewidth}
  \centerline{\includegraphics[scale=0.5]{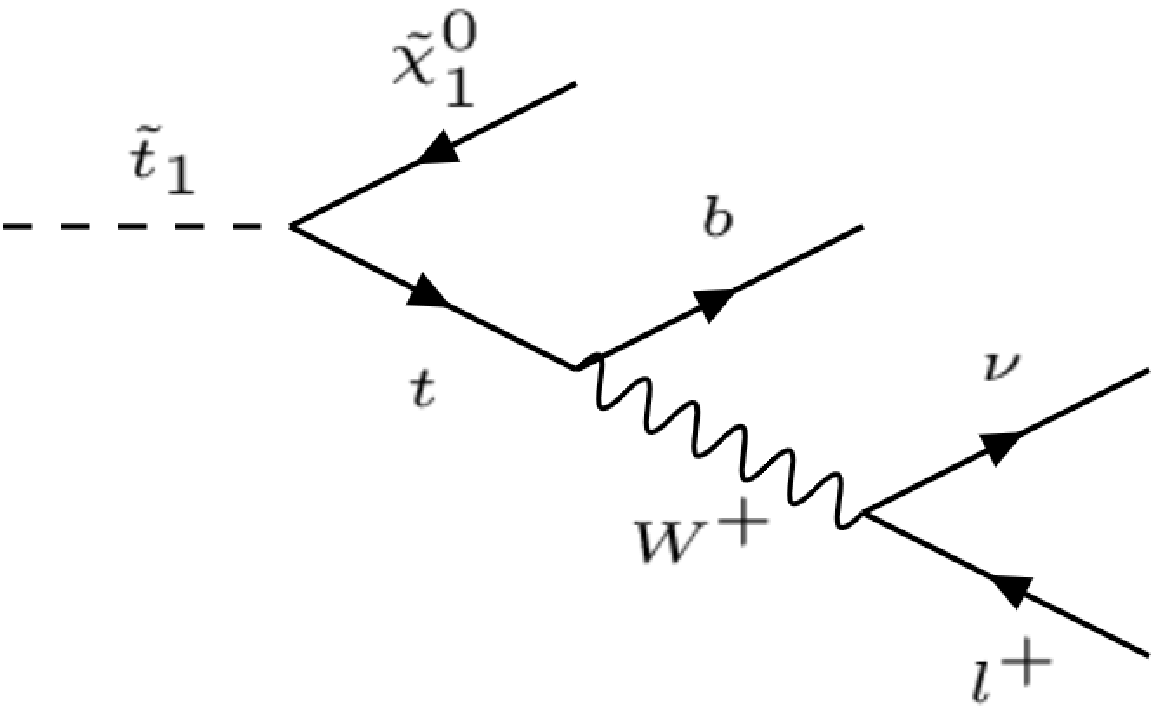}}
  \centerline{(a) Top channel of stop decay.}
\end{minipage}
\hfill
\begin{minipage}{0.48\linewidth}
  \centerline{\includegraphics[scale=0.5]{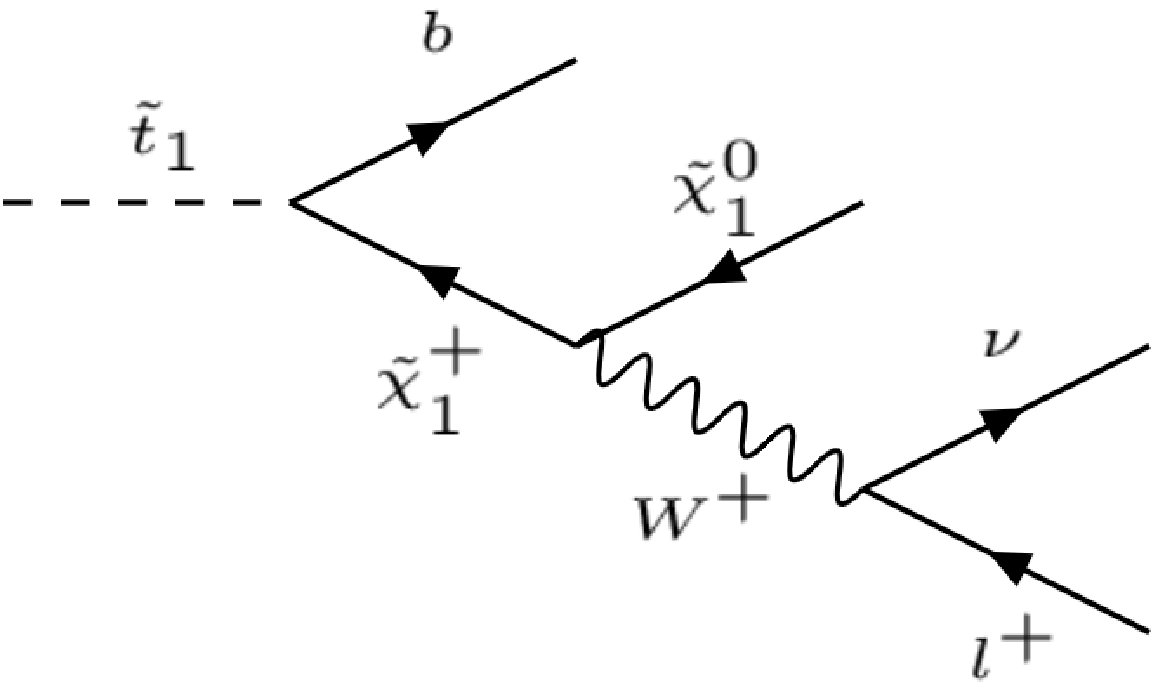}}
  \centerline{(b) Chargino channel of stop decay.}
\end{minipage}
\caption{Feynman diagrams of stop decays.}
\label{fig:FD}
\end{figure}

In the parent rest frame, the angular distribution of their decay product $f$ is given by,
\begin{eqnarray}
\frac{1}{\Gamma}\frac{d\Gamma}{d\cos\theta_{f}}=\frac{1}{2}(1+ \mathcal{P}_{f}\cos\theta_{f}),
\end{eqnarray}
where $\theta_{f}$ is the angle between the momentum of final state particle $f$ and the spin vector of parent in its rest frame and $\mathcal{P}_{f}$ is the spin analyzing power of final state particle $f$.
\begin{comment}
In order to measure the polarization, we can further define the forward-backward asymmetry $A_{fb}$ as:
\begin{eqnarray}
A_{fb}=\frac{\left(\int^{1}_{0}-\int^{0}_{-1}\right)d\cos\theta_{f}\cdot\frac{d\Gamma}{d\cos\theta_{f}}}{\left(\int^{1}_{0}+\int^{0}_{-1}\right)d\cos\theta_{f}\cdot\frac{d\Gamma}{d\cos\theta_{f}}}. \label{fba}
\end{eqnarray}
\end{comment}
$\mathcal{P}_{f}$ ranges from $-1$ to $1$ and reflects the degree of how much information about the parent top quark's spin the decay product $f$ carries. This quantity relates top spin to the decay product $f$'s angular distribution. In the following of this section, we will calculate the angular distributions of the charged lepton from the stop decays and its spin-analyzing power. Feynman diagrams relevant to the decay processes are shown in FIG.~\ref{fig:FD}. % The narrow width approximation are assumed in the calculations.

\subsection{$\tilde{t}_1 \to t \tilde{\chi}^0_1 \to (b W^+) \tilde{\chi}^0_1 \to (b \ell^+ \nu_\ell) \tilde{\chi}^0_1$}
We introduce the spin-projection operator $\hat{S}$, which is defined as
\begin{eqnarray}
\hat{S}=\frac{1}{2}(1+ \gamma^{5}\slashed s).
\end{eqnarray}
Here $s^{\mu}$ is the spin vector of the top quark. The matrix element of the process $\tilde{t}_{1}\rightarrow t\tilde{\chi}^{0}_{1}$ in the top rest frame is given by,
\begin{eqnarray}
|\mathcal{M}|^{2}=(g^{(t)}_{\textrm{eff}})^{2}m_{t}(E_{\tilde{\chi}^{0}_1}+m_{\tilde{\chi}^{0}_1}\sin2\theta^{(t)}_{\textrm{eff}}+\cos2\theta^{(t)}_{\textrm{eff}}\vec{p}_{\tilde{\chi}^{0}_1}\cdot\vec{s}),
\end{eqnarray}
where $E_{\tilde{\chi}^{0}}$ and $m_{\tilde{\chi}^{0}}$ denote the energy and mass of $\tilde{\chi}^{0}_1$, respectively. Then we can have the normalized differential decay width,
\begin{eqnarray}
\frac{1}{\Gamma}\frac{d\Gamma}{d\cos\theta_{t}}=\frac{1}{2}\left(1+\frac{|\vec{p}_{\tilde{\chi}^{0}}|\cos2\theta^{(t)}_{\textrm{eff}}}{E_{\tilde{\chi}^{0}}+m_{\tilde{\chi}^{0}}\sin2\theta^{(t)}_{\textrm{eff}}}\cos\theta_{t}\right),
\label{pt}
\end{eqnarray}
Since the direction of the charged lepton momentum in the decay of $t \to bl^{+}\nu$ is 100\% correlated with the top polarization at leading order~\cite{Jezabek:1994zv}, we can obtain the spin-analyzing power $\mathcal{P}^{(t)}_{l}$ for the charged lepton in the stop decay $\tilde{t}_{1} \to t\tilde{\chi}^{0}_{1} \to (W^+ b)\tilde{\chi}^{0}_{1} \to (l^{+}\nu b)\tilde{\chi}^{0}_{1}$,
\begin{eqnarray}
\mathcal{P}^{(t)}_{l}=\frac{|\vec{p}_{\tilde{\chi}^{0}}|\cos2\theta^{(t)}_{\textrm{eff}}}{E_{\tilde{\chi}^{0}}+m_{\tilde{\chi}^{0}}\sin2\theta^{(t)}_{\textrm{eff}}}.\label{plt}
\end{eqnarray}

\begin{figure}[t]
\centering
\includegraphics[scale=0.8]{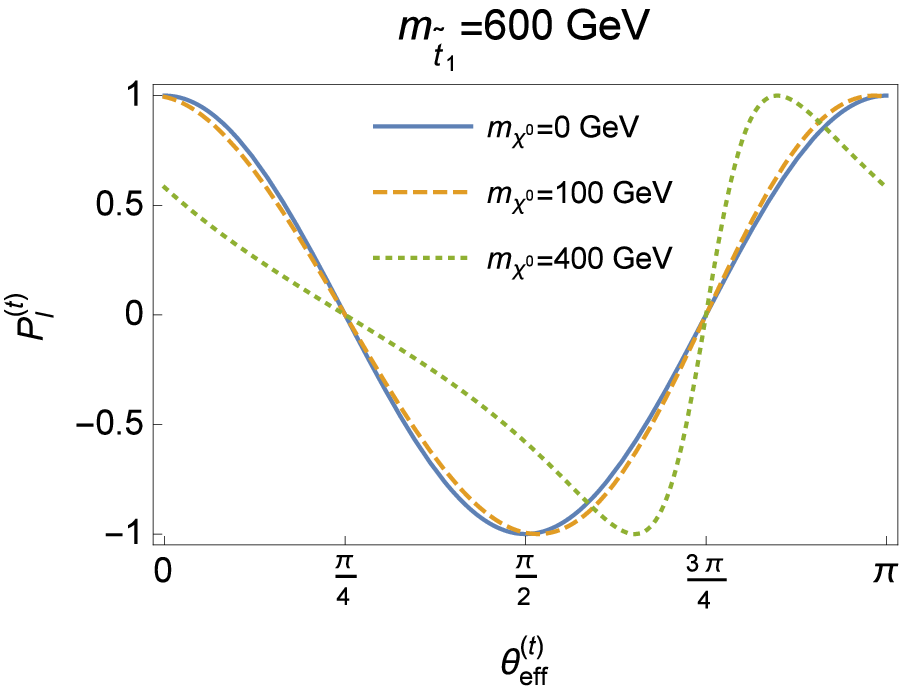}
\includegraphics[scale=0.8]{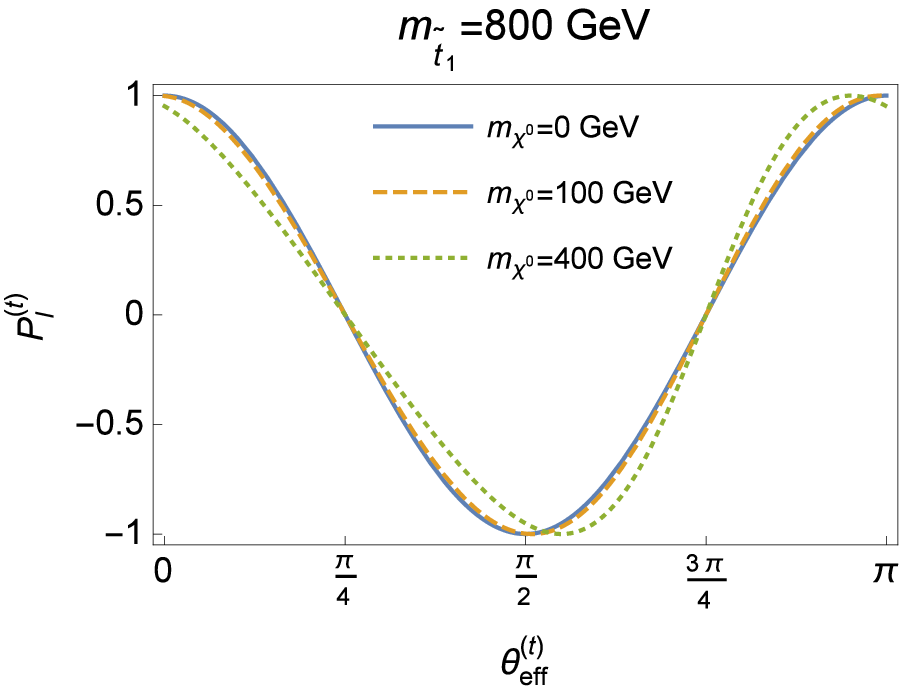}
\caption{Spin-analyzing power $\mathcal{P}^{(t)}_{l}$ for the charged lepton in the stop decay $\tilde{t}_{1} \to t\tilde{\chi}^{0}_{1} \to (W^+ b)\tilde{\chi}^{0}_{1} \to (l^{+}\nu b)\tilde{\chi}^{0}_{1}$.}
\label{fig:pt}
\end{figure}
In Fig.~\ref{fig:pt}, we show the $\mathcal{P}^{(t)}_{l}$ as a function of effective mixing angle $\theta^{(t)}_{\textrm{eff}}$ for different stop and neutralino masses. In the limit $m_{\tilde{\chi}^0_1}=0$, the distribution of $\mathcal{P}^{(t)}_{l}$ is symmetric around $\theta^{(t)}_{\textrm{eff}}=\pi/2$ and the top quark is fully left-handed (right-handed) for $\theta^{(t)}_{\textrm{eff}}=0~(\pi/2)$. While including the mass of $\tilde{\chi}^1_0$, the trend of the distribution will be changed. Besides, the degree of top polarization also depends on the mass difference between stop and neutralino. For a given massive neutralino, a heavier stop will make $\mathcal{P}^{(t)}_{l}$ distribution closer to that for a massless neutralino. In Tab.~\ref{tab:pt}, we present the numerical values of $\mathcal{P}^{(t)}_{l}$ for several benchmark points. Recent searches for top squark already placed exclusion limits~\cite{Sirunyan:2017leh} on $\tilde{t}_{1}$ mass $< 1050$ GeV with $\tilde{\chi}^{0}_{1}$ mass $< 500$ GeV, but it should also be noted that these limits are given by assuming a 100\% branching ratio of stop decay into $t\tilde{\chi}^{0}_{1}$ which can be relaxed in a specific model~\cite{Han:2016xet}, and similarly for the case in Tab.~\ref{tab:ptchi} below. So the benchmarks shown in Tab.~\ref{tab:pt} and \ref{tab:ptchi} are still allowed.

\begin{table}
\begin{tabular}{c|c|c|c|c|c||c|c|c|c|c}
\hline\hline
top-channel & \multicolumn{5}{c||}{\,$m_{\tilde{t}_{1}}=600\textrm{GeV},\,m_{\tilde{\chi}^{0}_{1}}=0\textrm{GeV}$\,} & \multicolumn{5}{c}{\,$m_{\tilde{t}_{1}}=800\textrm{GeV},\,m_{\tilde{\chi}^{0}_{1}}=0\textrm{GeV}$\,} \\  \hline
 $\theta^{(t)}_{\textrm{eff}}$ & $0$ & $\frac{\pi}{8}$ & $\frac{\pi}{2}$ & $\frac{5\pi}{8}$ & $\pi$ & $0$ & $\frac{\pi}{8}$ & $\frac{\pi}{2}$ & $\frac{5\pi}{8}$ & $\pi$ \\  \hline
$\mathcal{P}^{(t)}_{l}$ & 1.0 & 0.7 & -1.0 & -0.7 & 1.0 & 1.0 & 0.7 & -1.0 & -0.7 & 1.0 \\ \hline\hline
top-channel & \multicolumn{5}{c||}{$m_{\tilde{t}_{1}}=600\textrm{GeV},\,m_{\tilde{\chi}^{0}_{1}}=100\textrm{GeV}$} & \multicolumn{5}{c}{$m_{\tilde{t}_{1}}=800\textrm{GeV},\,m_{\tilde{\chi}^{0}_{1}}=100\textrm{GeV}$} \\ \hline
$\theta^{(t)}_{\textrm{eff}}$ & $0$ & $\frac{\pi}{8}$ & $\frac{\pi}{2}$ & $\frac{5\pi}{8}$ & $\pi$ & $0$ & $\frac{\pi}{8}$ & $\frac{\pi}{2}$ & $\frac{5\pi}{8}$ & $\pi$ \\  \hline
$\mathcal{P}^{(t)}_{l}$ & 0.99 & 0.65 & -0.99 & -0.76 & 0.99 & 0.99 & 0.67 & -0.99 & -0.73 & 0.99 \\ \hline\hline
top-channel & \multicolumn{5}{c||}{$m_{\tilde{t}_{1}}=600\textrm{GeV},\,m_{\tilde{\chi}^{0}_{1}}=400\textrm{GeV}$} & \multicolumn{5}{c}{$m_{\tilde{t}_{1}}=800\textrm{GeV},\,m_{\tilde{\chi}^{0}_{1}}=400\textrm{GeV}$} \\ \hline
$\theta^{(t)}_{\textrm{eff}}$ & $0$ & $\frac{\pi}{8}$ & $\frac{\pi}{2}$ & $\frac{5\pi}{8}$ & $\pi$ & $0$ & $\frac{\pi}{8}$ & $\frac{\pi}{2}$ & $\frac{5\pi}{8}$ & $\pi$ \\  \hline
$\mathcal{P}^{(t)}_{l}$ & 0.58 & 0.26 & -0.58 & -0.96 & 0.58 & 0.95 & 0.55 & -0.95 & -0.86 & 0.95 \\ \hline\hline
\end{tabular}
\caption{The values of $\mathcal{P}^{(t)}_{l}$ for benchmark points in the stop decay $\tilde{t}_{1} \to t\tilde{\chi}^{0}_{1} \to (W^+ b)\tilde{\chi}^{0}_{1} \to (l^{+}\nu b)\tilde{\chi}^{0}_{1}$.}
\label{tab:pt}
\end{table}

\subsection{$\tilde{t}_1 \to b\tilde{\chi}^+_1 \to b (W^+ \tilde{\chi}^0_1) \to b (\ell^+ \nu \tilde{\chi}^0_1)$}
Similarly, the normalized differential decay width of $\tilde{t}_1 \to b\tilde{\chi}^+_1$ is given by,
\begin{eqnarray}
\frac{1}{\Gamma}\frac{d\Gamma}{d\cos\theta_{\tilde{\chi}^+_1}}=\frac{1}{2}\left(1+\frac{|\vec{p}_{b}|\cos2\theta^{(\chi)}_{\textrm{eff}}}{E_{b}-m_{b}\sin2\theta^{(\chi)}_{\textrm{eff}}}\cos\theta_{\tilde{\chi}^+_1}\right).
\label{pchi}
\end{eqnarray}
Then, we calculate the matrix element of the process $\tilde{\chi}^+_1 \to  W^+ \tilde{\chi}^0_1 \to  \ell^+ \nu \tilde{\chi}^0_1 $,
%\begin{eqnarray}
%|\mathcal{M}|^{2}=\frac{(g^{(W)}_{\textrm{eff}})^{2}(g/2)^{2}}{(p^{2}_{W}-m^{2}_{W})^{2}+m^{2}_{W}\Gamma^{2}_{W}}\sum_{k,k'=L,R}W_{\mu\nu}T^{\mu\nu}_{kk'}\,,
%\end{eqnarray}
%with
%\begin{eqnarray}
%W_{\mu\nu}={}&P_{\mu\rho}P_{\nu\sigma}\textrm{Tr}\left[\gamma^{\rho}p\!\!\!/_{l}\gamma^{\sigma}P_{L}p\!\!\!/_{\nu}\right],\\
%P^{\mu\rho}={}&-g^{\mu\rho}+\frac{p^{\mu}_{W}p^{\rho}_{W}}{m^{2}_{W}},\\
%T^{\mu\nu}_{kk'}={}&c_{k}c_{k'}\textrm{Tr}\left[\gamma^{\mu}P_{k}\hat{S}(p\!\!\!/_{\tilde{\chi}^{+}_{1}}+m_{\tilde{\chi}^{+}_{1}})\gamma^{\nu}P_{k'}(p\!\!\!/_{\tilde{\chi}^{0}_{1}}+m_{\tilde{\chi}^{0}_{1}})\right],\quad %k,k'=L,R,
%\end{eqnarray}
\begin{eqnarray}
|\mathcal{M}|^{2}
=\frac{(g^{(W)}_{\textrm{eff}})^{2}g^{2}_2}{2m^{2}_{W}\Gamma^{2}_{W}}\Biggl\{8m^{2}_{\tilde{\chi}^{+}_{1}}E_{l}E_{\nu}-4m_{\tilde{\chi}^{+}_{1}}m^{2}_{W}(c^{2}_{L}E_{l}+c^{2}_{R}E_{\nu}+c_{L}c_{R}m_{\tilde{\chi}^{0}_{1}})\nonumber\\
+4m_{\tilde{\chi}^{+}_{1}}|\vec{s}|\cos\theta_{l}\Bigl[c_{L}E_{l}(2c_{L}m_{\tilde{\chi}^{+}_{1}}E_{\nu}-c_{L}m^{2}_{W}-2c_{R}m_{\tilde{\chi}^{0}_{1}}E_{\nu})\nonumber\\
+c_{R}\left(\frac{m^{2}_{W}}{2E_{l}}-E_{\nu}\right)(2c_{R}m_{\tilde{\chi}^{+}_{1}}E_{l}-c_{R}m^{2}_{W}-2c_{L}m_{\tilde{\chi}^{0}_{1}}E_{l})\Bigr]\Biggr\},
\end{eqnarray}
where $g_2$ is the $SU(2)_{L}$ coupling, $c_{L}=\sin\theta^{(W)}_{\textrm{eff}}$ and $c_{R}=\cos\theta^{(W)}_{\textrm{eff}}$. $\Gamma_{W}$ is the decay width of $W$ boson. An on-shell $W$ boson and narrow width approximation are assumed in the calculation. The normalized angular distribution of the charged lepton in the process $\tilde{\chi}^+_1 \to  W^+ \tilde{\chi}^0_1 \to  \ell^+ \nu \tilde{\chi}^0_1 $ is given by,
\begin{eqnarray}
\frac{1}{\Gamma}\frac{d\Gamma}{d\cos\theta_{l}}=\frac{1}{2}\left(1+\frac{S_{2}(m_{\tilde{\chi}^{+}_{1}},m_{\tilde{\chi}^{0}_{1}},m_{W})}{S_{1}(m_{\tilde{\chi}^{+}_{1}},m_{\tilde{\chi}^{0}_{1}},m_{W})}\cos\theta_{l}\right),
\label{pt_chi}
\end{eqnarray}
with
\begin{eqnarray}
S_{1}(m_{\tilde{\chi}^{+}_{1}},m_{\tilde{\chi}^{0}_{1}},m_{W})&=&\frac{3\lambda^{2}}{16m^{3}_{\tilde{\chi}^{+}_{1}}}+\frac{5\lambda m^{2}_{W}E_{W}}{4m^{2}_{\tilde{\chi}^{+}_{1}}}-\frac{5\lambda E^{2}_{W}}{2m_{\tilde{\chi}^{+}_{1}}}-15E^{3}_{W}m^{2}_{W}+15m_{\tilde{\chi}^{+}_{1}}E^{4}_{W}\nonumber\\
&&+10c_{L}c_{R}m^{2}_{W}m_{\tilde{\chi}^{0}_{1}}\left(\frac{\lambda}{4m^{2}_{\tilde{\chi}^{+}_{1}}}-3E^{2}_{W}\right),\\
S_{2}(m_{\tilde{\chi}^{+}_{1}},m_{\tilde{\chi}^{0}_{1}},m_{W})&=&(c^{2}_{L}-c^{2}_{R})\left(\frac{3\lambda^{2}}{16m^{3}_{\tilde{\chi}^{+}_{1}}}+\frac{5\lambda m^{2}_{W}E_{W}}{4m^{2}_{\tilde{\chi}^{+}_{1}}}-\frac{5\lambda E^{2}_{W}}{2m_{\tilde{\chi}^{+}_{1}}}-15E^{3}_{W}m^{2}_{W}+15m_{\tilde{\chi}^{+}_{1}}E^{4}_{W}\right)\nonumber\\
&&-c^{2}_{R}\left(\frac{5\lambda m^{2}_{W}}{2m_{\tilde{\chi}^{+}_{1}}}+30m^{4}_{W}E_{W}-30m_{\tilde{\chi}^{+}_{1}}m^{2}_{W}E^{2}_{W}\right) \nonumber\\
&&+10c_{L}c_{R}m^{2}_{W}m_{\tilde{\chi}^{0}_{1}}\left(\frac{\lambda}{4m^{2}_{\tilde{\chi}^{+}_{1}}}-3E^{2}_{W}\right),
\end{eqnarray}
where the kinematical function $\lambda$ is defined as,
\begin{eqnarray}
\lambda(x,y,z)\equiv x^{2}+y^{2}+z^{2}-2xy-2yz-2xz,
\end{eqnarray}
with $x=m^2_{\tilde{\chi}^{+}_{1}}$, $y=m^2_{\tilde{\chi}^{0}_{1}}$ and $z=m^2_{W}$. Therefore, we can have the spin-analyzing power of the charged lepton in the chargino rest frame,
\begin{eqnarray}
\mathcal{P}^{(\chi)}_{l}=\frac{|\vec{p}_{b}|\cos2\theta^{(\chi)}_{\textrm{eff}}}{E_{b}-m_{b}\sin2\theta^{(\chi)}_{\textrm{eff}}}\times\frac{S_{2}(m_{\tilde{\chi}^{+}_{1}},m_{\tilde{\chi}^{0}_{1}},m_{W})}{S_{1}(m_{\tilde{\chi}^{+}_{1}},m_{\tilde{\chi}^{0}_{1}},m_{W})} \label{plchi}
%\\{}&\xrightarrow{m_{b}\rightarrow0}\cos2\theta^{(\chi)}_{\textrm{eff}}\times\frac{S_{2}(m_{\tilde{\chi}^{+}_{1}},m_{\tilde{\chi}^{0}_{1}},m_{W})}{S_{1}(m_{\tilde{\chi}^{+}_{1}},m_{\tilde{\chi}^{0}_{1}},m_{W})},\label{plchilim}
\end{eqnarray}

\begin{figure}[t]
\centering
\includegraphics[scale=0.8]{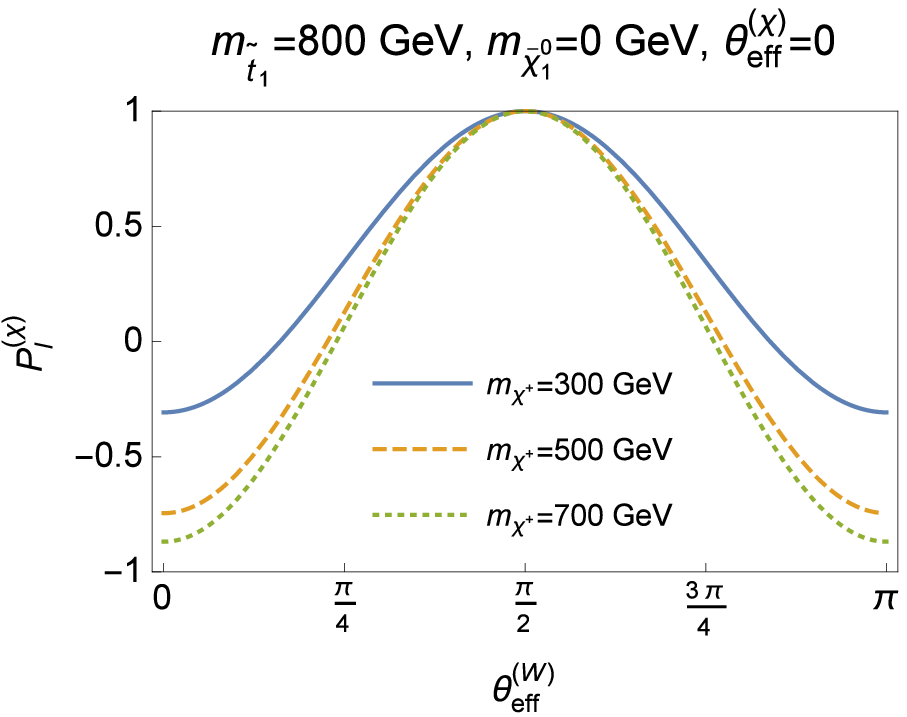}
\includegraphics[scale=0.8]{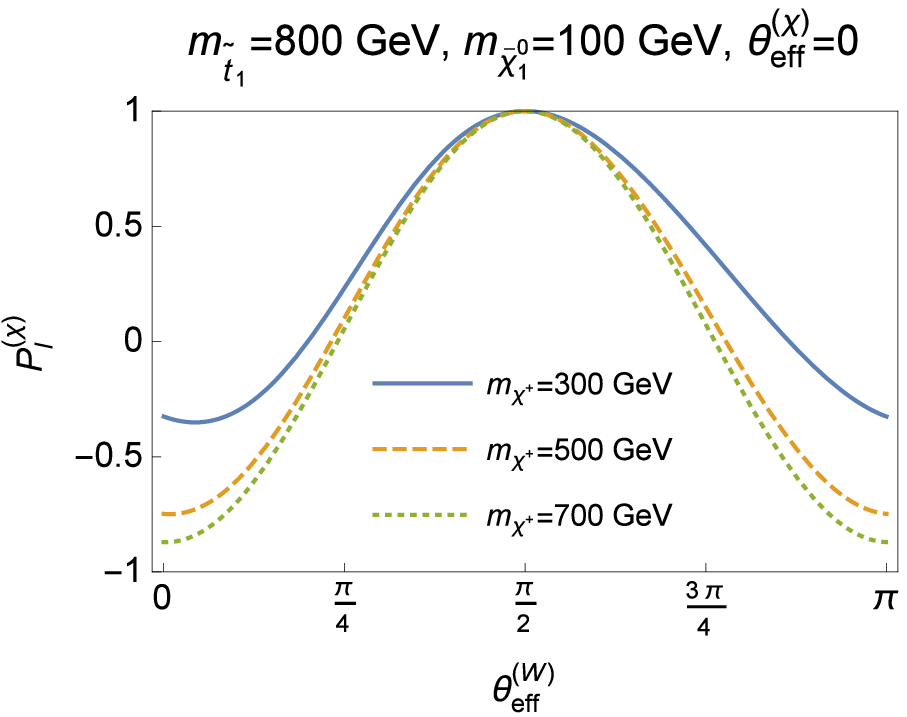}
\caption{Spin-analyzing power $\mathcal{P}^{(\chi)}_{l}$ for the charged lepton in the stop decay $\tilde{t}_1 \to b\tilde{\chi}^+_1 \to b (W^+ \tilde{\chi}^0_1) \to b (\ell^+ \nu \tilde{\chi}^0_1)$.}
\label{fig:pl_chi}
\end{figure}
In Fig.~\ref{fig:pl_chi}, we plot the spin-analyzing power of the charged lepton in the chargino decay channel as a function of $\theta^{(W)}_{\textrm{eff}}$. It should be mentioned that $\mathcal{P}^{(\chi)}_{l}$ will be independent of stop mass in the limit of $m_b=0$, and $\mathcal{P}^{(\chi)}_{l}$ will become $-\mathcal{P}^{(\chi)}_{l}$ when $\theta^{(\chi)}_{\textrm{eff}} \to \theta^{(\chi)}_{\textrm{eff}}+\frac{\pi}{2}$. Thus, we take $\theta^{(\chi)}_{\textrm{eff}}=0$ in Fig.~\ref{fig:pl_chi} for example, which indicates that the chargino is fully left-handed polarized. It can be seen that $\mathcal{P}^{(\chi)}_{l}$ can reach the maximal value of spin-analyzing power at $\theta^{(W)}_{\textrm{eff}}=\pi/2$, due to equal $S_{1}$ and $S_{2}$. In Tab.~\ref{tab:ptchi}, we present the numerical values of $\mathcal{P}^{(\chi)}_{l}$ for several benchmark points.

\begin{table}
\begin{tabular}{c|c|c|c|c|c||c|c|c|c|c}
\hline\hline
$\tilde{\chi}^+_1$-channel & \multicolumn{5}{c||}{\,$m_{\tilde{\chi}^{0}_{1}}=0\textrm{GeV},\,m_{\tilde{\chi}^{+}_{1}}=300\textrm{GeV}$\,} & \multicolumn{5}{c}{\,$m_{\tilde{\chi}^{0}_{1}}=100~\textrm{GeV},\,m_{\tilde{\chi}^{+}_{1}}=300~\textrm{GeV}$\,} \\  \hline
 $\theta^{(W)}_{\textrm{eff}}$ & $0$ & $\frac{\pi}{4}$ & $\frac{\pi}{2}$ & $\frac{3\pi}{4}$ & $\pi$ & $0$ & $\frac{\pi}{4}$ & $\frac{\pi}{2}$ & $\frac{3\pi}{4}$ & $\pi$ \\  \hline
$\mathcal{P}^{(\chi)}_{l}$ & -0.31 & \,0.34\, & \,\,\,\,1.0\,\,\,\, & \,0.34\, & -0.31 & -0.38 & \,0.23\, & \,\,\,\,1.0\,\,\,\, & \,0.41\, & -0.32 \\ \hline\hline
$\tilde{\chi}^+_1$-channel & \multicolumn{5}{c||}{$m_{\tilde{\chi}^{0}_{1}}=0\textrm{GeV},\,m_{\tilde{\chi}^{+}_{1}}=500\textrm{GeV}$} & \multicolumn{5}{c}{$m_{\tilde{\chi}^{0}_{1}}=100\textrm{GeV},\,m_{\tilde{\chi}^{+}_{1}}=500\textrm{GeV}$} \\ \hline
$\theta^{(W)}_{\textrm{eff}}$ & $0$ & $\frac{\pi}{4}$ & $\frac{\pi}{2}$ & $\frac{3\pi}{4}$ & $\pi$ & $0$ & $\frac{\pi}{4}$ & $\frac{\pi}{2}$ & $\frac{3\pi}{4}$ & $\pi$ \\  \hline
$\mathcal{P}^{(\chi)}_{l}$ & -0.74 & 0.12 & 1.0 & 0.12 & -0.74 & -0.74 & 0.10 & 1.0 & 0.15 & -0.74 \\ \hline\hline
$\tilde{\chi}^+_1$-channel & \multicolumn{5}{c||}{$m_{\tilde{\chi}^{0}_{1}}=0\textrm{GeV},\,m_{\tilde{\chi}^{+}_{1}}=700\textrm{GeV}$} & \multicolumn{5}{c}{$m_{\tilde{\chi}^{0}_{1}}=100\textrm{GeV},\,m_{\tilde{\chi}^{+}_{1}}=700\textrm{GeV}$} \\ \hline
$\theta^{(W)}_{\textrm{eff}}$ & $0$ & $\frac{\pi}{4}$ & $\frac{\pi}{2}$ & $\frac{3\pi}{4}$ & $\pi$ & $0$ & $\frac{\pi}{4}$ & $\frac{\pi}{2}$ & $\frac{3\pi}{4}$ & $\pi$ \\  \hline
$\mathcal{P}^{(\chi)}_{l}$ & -0.87 & 0.06 & 1.0 & 0.06 & -0.87 & -0.87 & 0.05 & 1.0 & 0.07 & -0.87 \\ \hline\hline
\end{tabular}
\caption{The values of $\mathcal{P}^{(\chi)}_{l}$ for benchmark points in the stop decay $\tilde{t}_{1} \to b\tilde{\chi}^{+}_{1} \to b(W^+\tilde{\chi}^{0}_{1}) \to b(l^{+}\nu\tilde{\chi}^{0}_{1})$.}
\label{tab:ptchi}
\end{table}
%In the limit $\cos\theta^{(W)}_{\textrm{eff}}=0$, $S_{1}$ equals $S_{2}$ and hence the case like the top decay in which the decayed charged lepton has the maximal spin-analyzing power.

%which leads to a different expression of normalized doubly differential spectra in the chargino rest frame
%\begin{align}
%\frac{1}{\Gamma}\frac{d\Gamma}{dE_{l}d\cos\theta_{l}}=\frac{1}{2^{6}\pi^{3}}\frac{E_{\nu}E_{l}}{m_{\tilde{\chi}^{+}_{1}}E_{\tilde{\chi}^{0}_{1}}}\frac{|\mathcal{M}|^{2}}{\Gamma},
%\end{align}
%in which the chargino's decay width $\Gamma$ is
%\begin{align}
%\Gamma=\frac{(g^{(W)}_{\textrm{eff}})^{2}g^{2}\lambda^{1/2}}{2^{6}\pi^{3}(m^{2}_{\tilde{\chi}^{+}_{1}}-m^{2}_{W})m^{2}_{W}\Gamma^{2}_{W}}\Biggl[{}&\frac{\lambda^{2}}{160m^{3}_{\tilde{\chi}^{+}_{1}}}+\frac{\lambda m^{2}_{W}E_{W}}{24m^{2}_{\tilde{\chi}^{+}_{1}}}-\frac{\lambda E^{2}_{W}}{12m_{\tilde{\chi}^{+}_{1}}}-\frac{E^{3}_{W}}{2}\left(\frac{\lambda}{16m^{2}_{\tilde{\chi}^{+}_{1}}}-m_{\tilde{\chi}^{+}_{1}}E_{W}\right)\notag\\
%{}&+c_{L}c_{R}m^{2}_{W}m_{\tilde{\chi}^{0}_{1}}\left(\frac{\lambda}{12m^{2}_{\tilde{\chi}^{+}_{1}}}-E^{2}_{W}\right)\Biggr],
%\end{align}
%and $\lambda=\lambda(m^{2}_{\tilde{\chi}^{+}_{1}},m^{2}_{\tilde{\chi}^{0}_{1}},m^{2}_{W})$ is defined as
%\begin{align}
%\lambda(x,y,z)\equiv x^{2}+y^{2}+z^{2}-2xy-2yz-2xz.
%\end{align}

%\begin{align}
%E_{\nu}=E_{W}-E_{l}=\frac{1}{2}\left(m_{\tilde{\chi}^{+}_{1}}+\frac{m^{2}_{W}-m^{2}_{\tilde{\chi}^{0}_{1}}}{m_{\tilde{\chi}^{+}_{1}}}\right)-E_{l}.
%\end{align}

\section{Top polarization effects in NSUSY}
\begin{figure}[t]
\centering
\includegraphics[height=4cm,width=4cm]{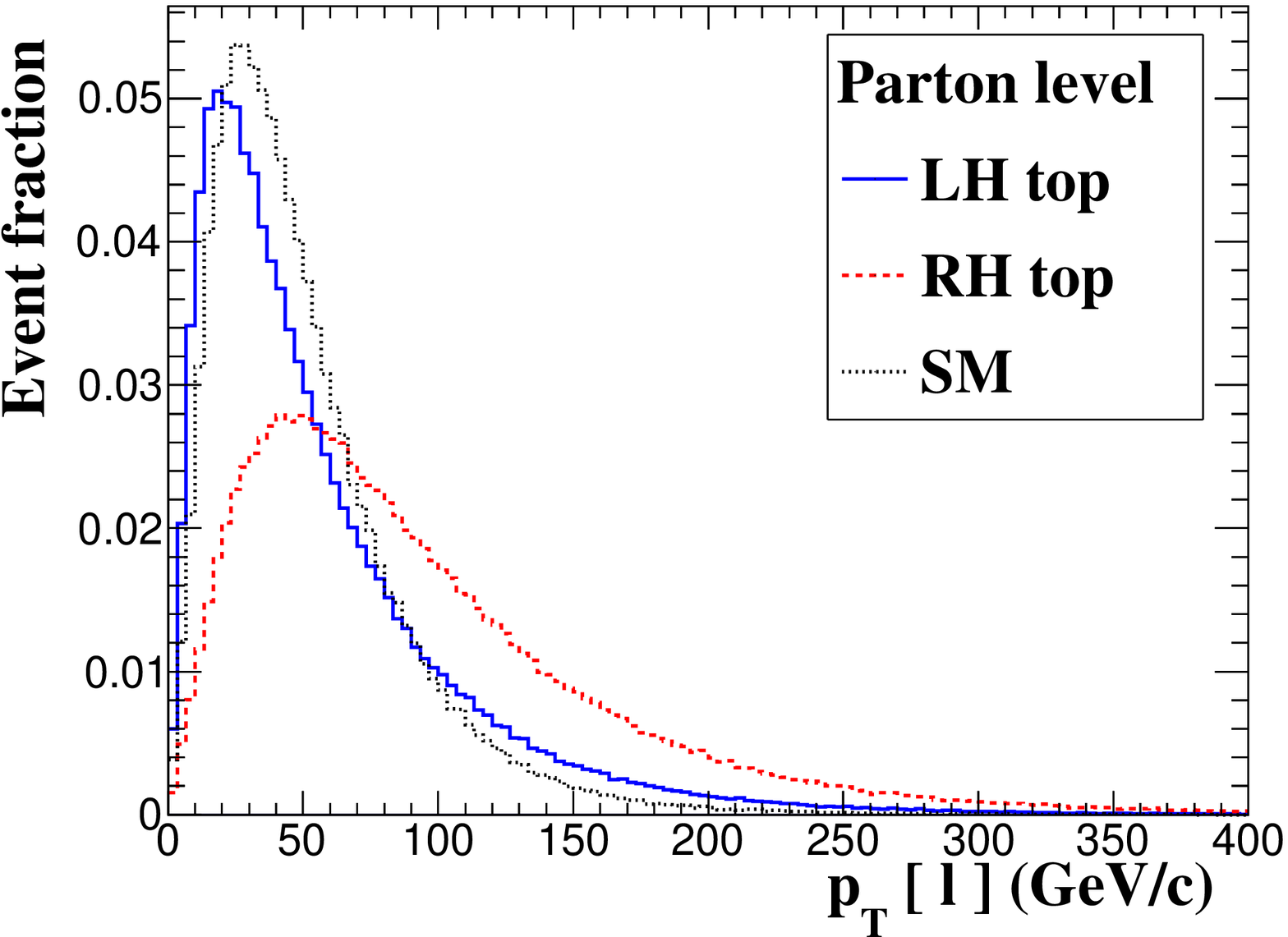}
\includegraphics[height=4cm,width=4cm]{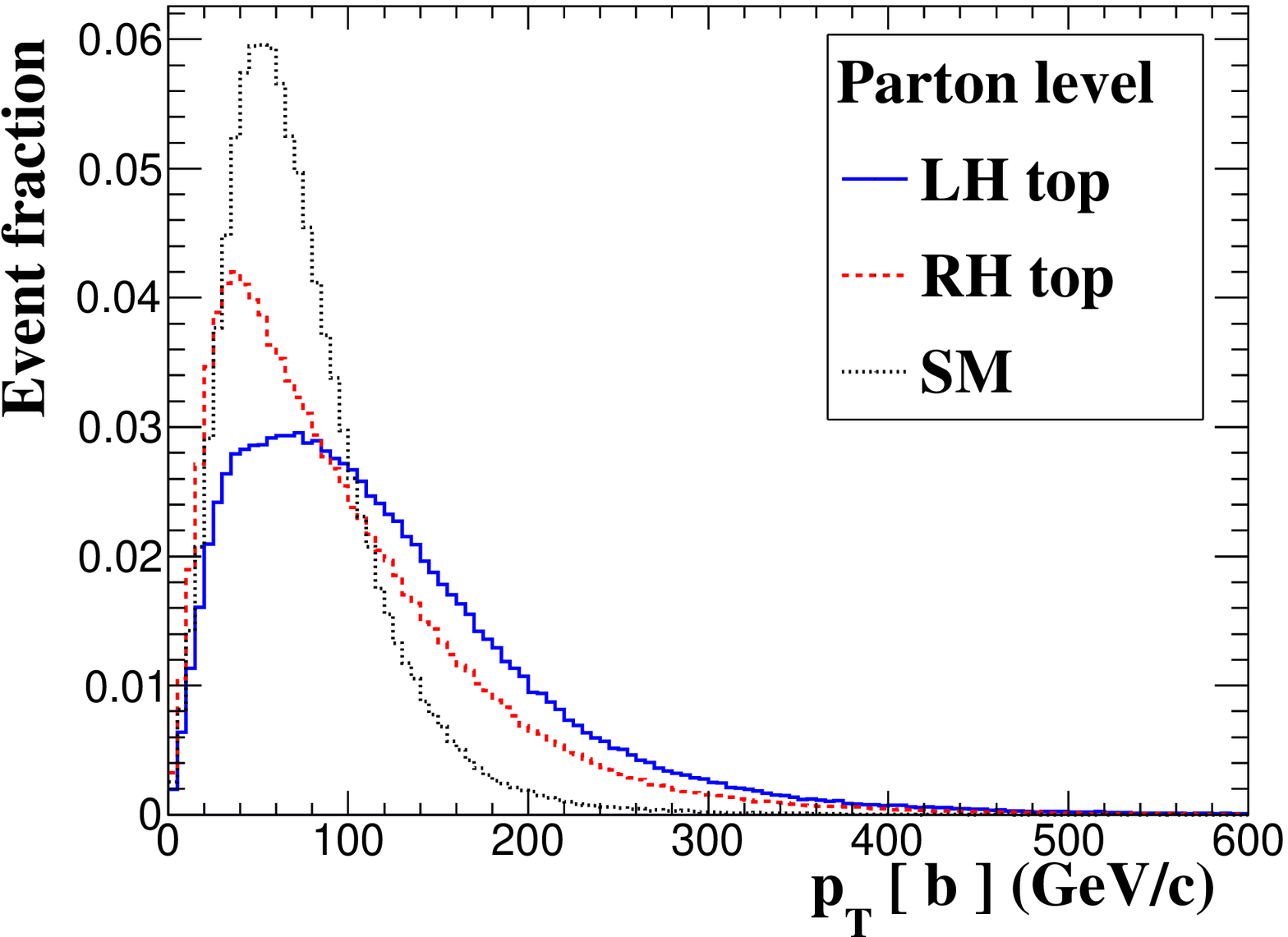}
\includegraphics[height=4cm,width=4cm]{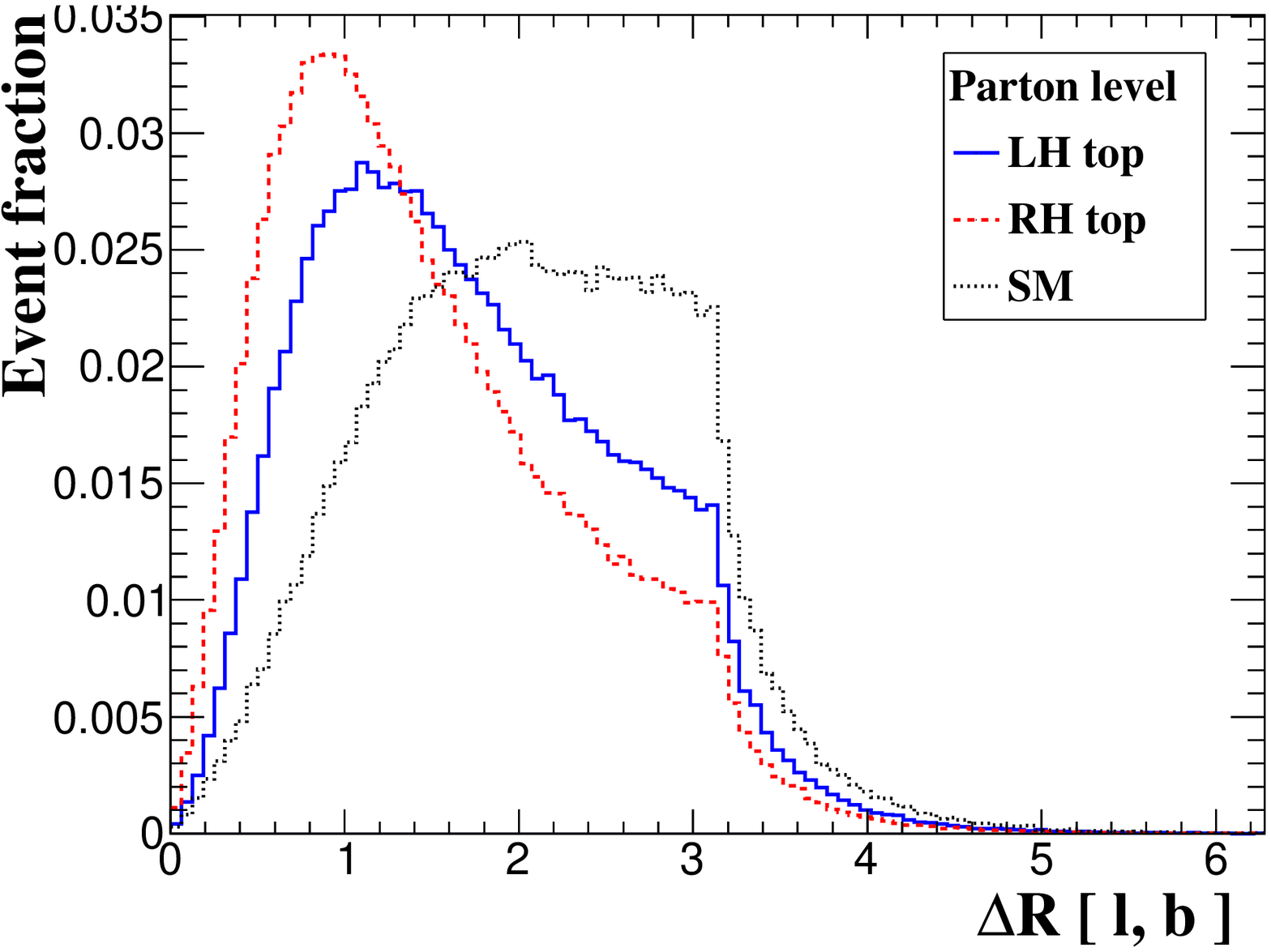}
\includegraphics[height=4cm,width=4cm]{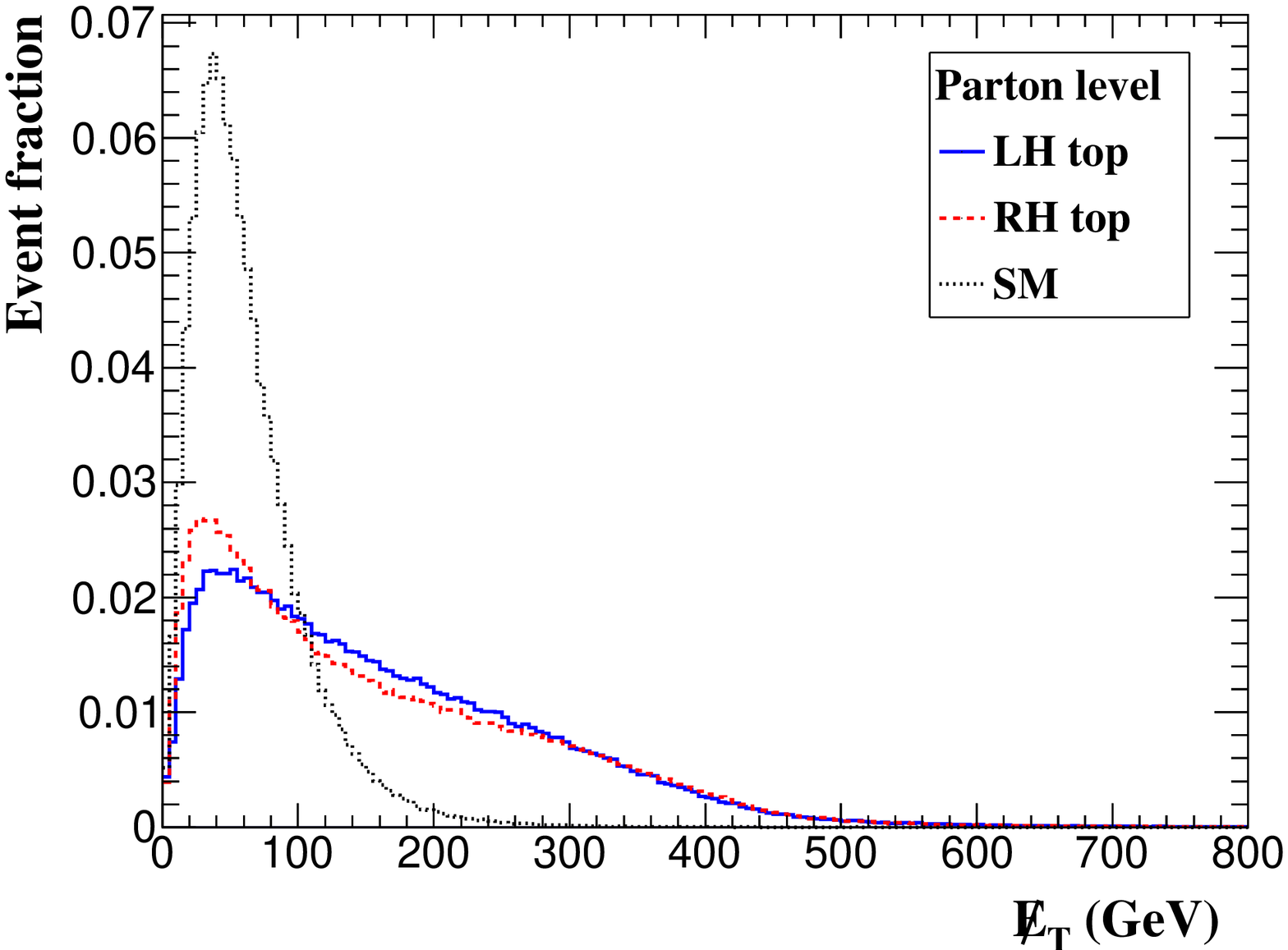}\\[7pt]
\includegraphics[height=4cm,width=4cm]{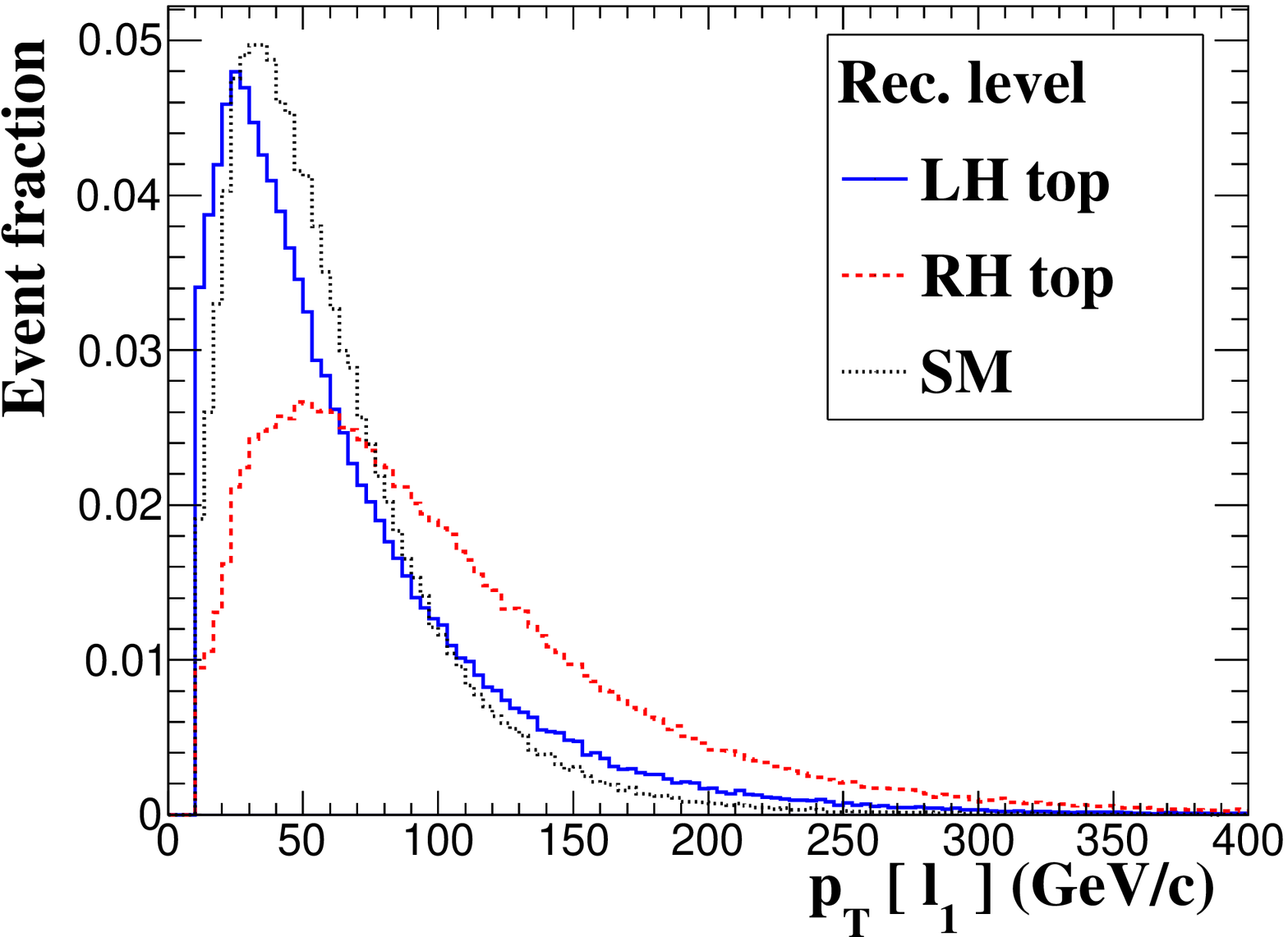}
\includegraphics[height=4cm,width=4cm]{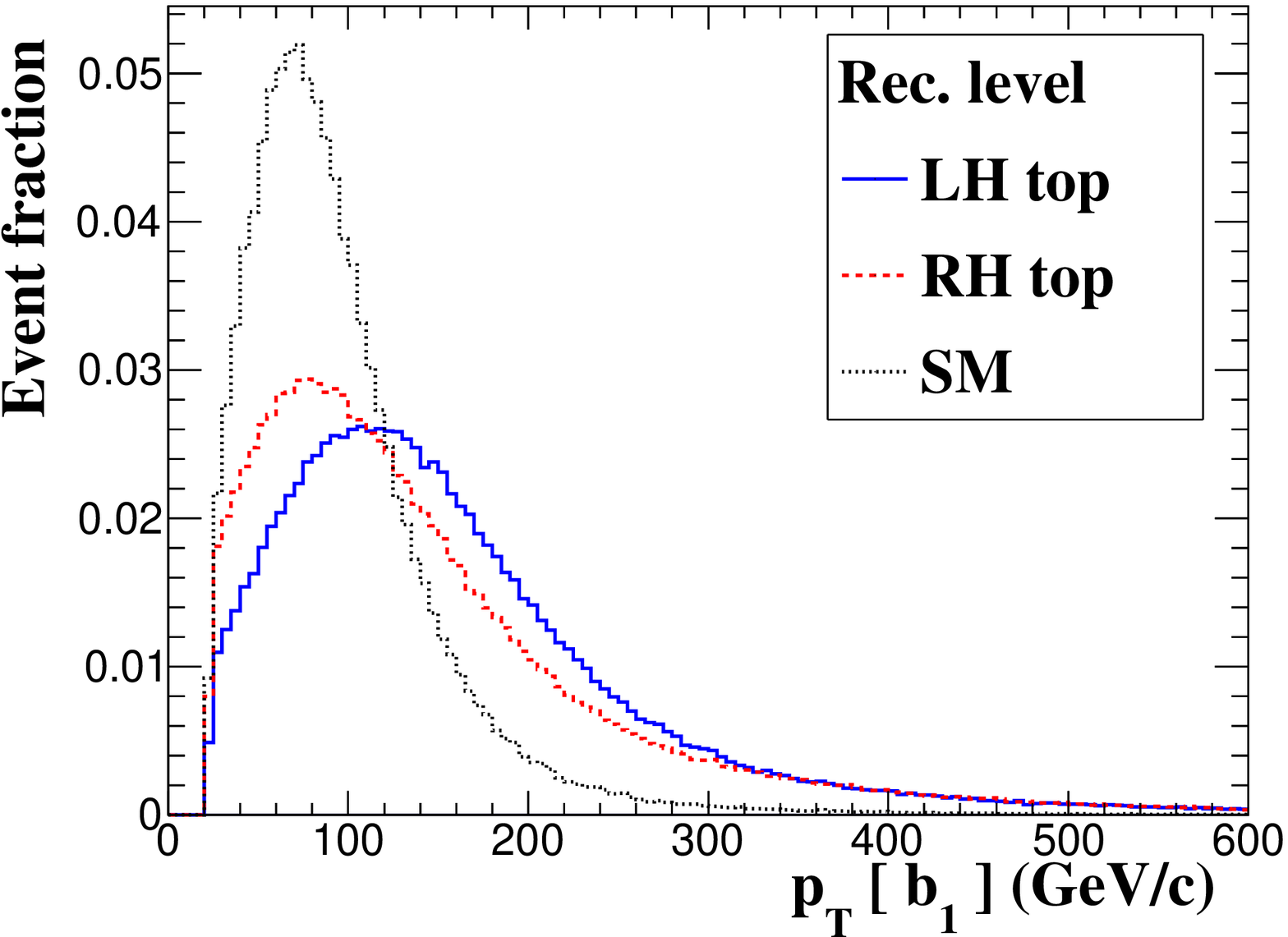}
\includegraphics[height=4cm,width=4cm]{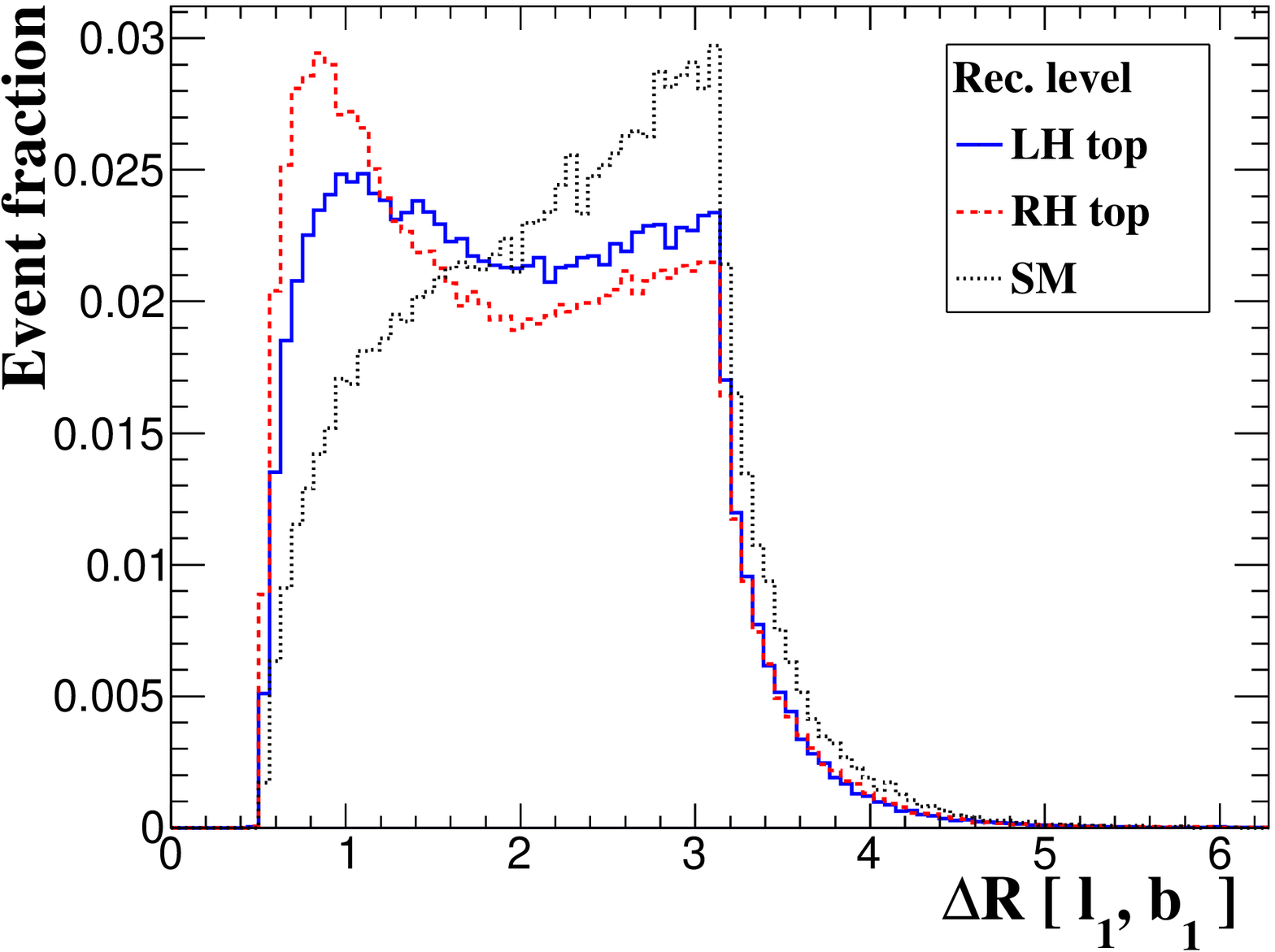}
\includegraphics[height=4cm,width=4cm]{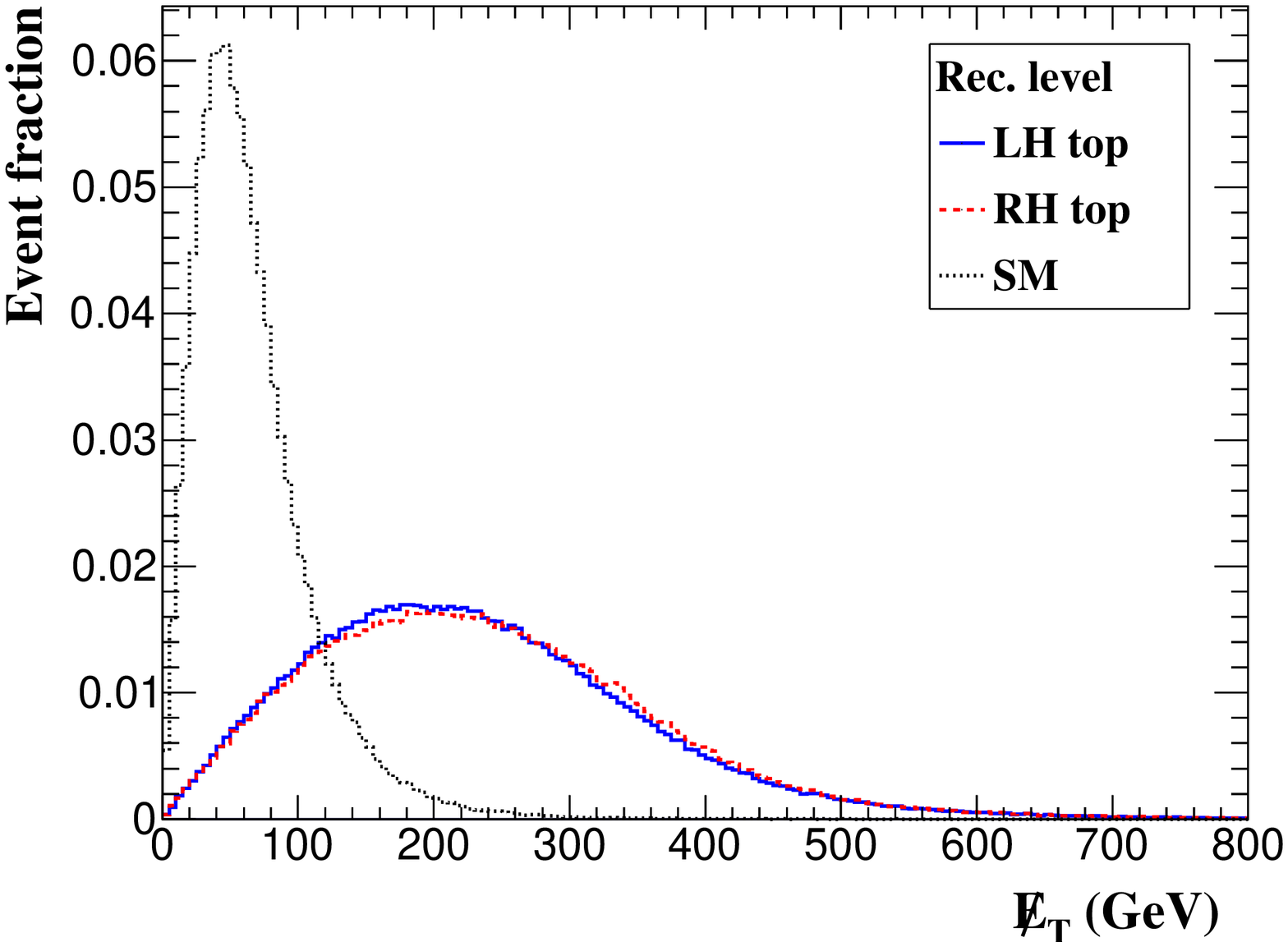}
\includegraphics[height=4cm,width=4cm]{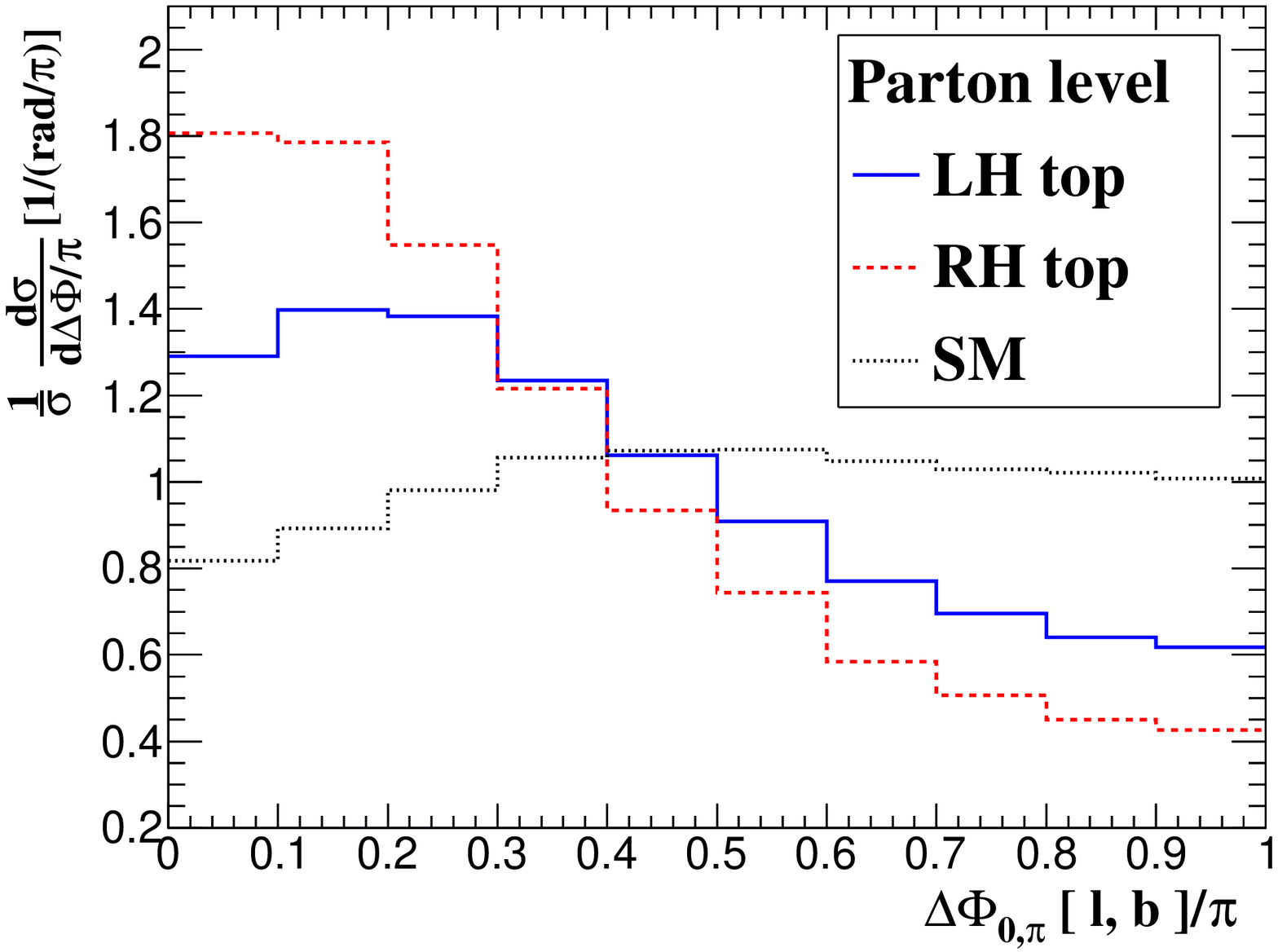}
\includegraphics[height=4cm,width=4cm]{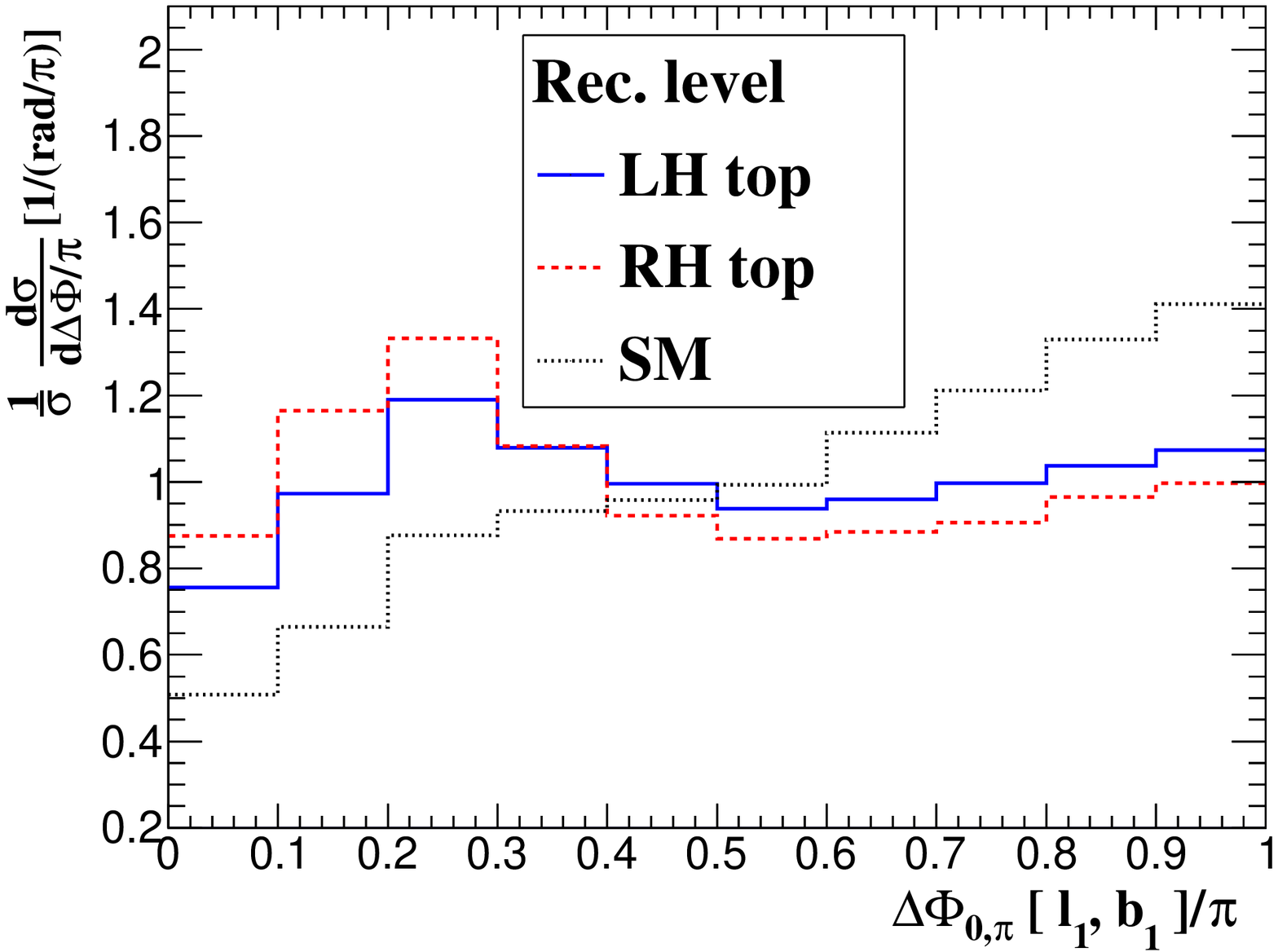}
\caption{Kinematical distributions of stop semileptonic channel $pp \to \tilde{t}_1 \tilde{t}^*_1 \to t (\to b j j) \bar{t} (\to \bar{b} \ell^- \bar{\nu}) \tilde{\chi}^0_{1,2} \tilde{\chi}^0_{1,2}$ and $t\bar{t}$ background events at parton- and reconstruction-level at 13 TeV LHC.}
\label{fig:semi_spin}
\end{figure}
Among various SUSY models, the natural SUSY is a well-motivated scenario~\cite{Hall:2011aa,Papucci:2011wy}. In NSUSY, only a small set of sparticles that closely relate to the naturalness are involved, such as higgsinos, stops and gluinos. The phenomenology of stop has been studied in~\cite{Baer:2018hpb,Baer:2017pba,Baer:2016bwh,Han:2016xet,Kobakhidze:2015scd,Kobakhidze:2015dra}. Other irrelevant sparticles are assumed to be heavy and decoupled from the spectrum of NSUSY. When $\mu \ll M_{1,2}$, the electroweakino mixing matrices $N_{ij}$, $U_{ij}$ and $V_{ij}$ can be given by,
\begin{align}
N_{11,12,21,22},V_{11},U_{11}\sim0,\,\,N_{13,14,23}=-N_{24}\sim\frac{\sqrt{2}}{2},\,\,V_{12}\sim \textrm{sgn}(\mu),\,\,U_{12}\sim1. \label{higgsino_param}
\end{align}
This leads to the fact that the polarization of the top quark and chargino from stop decays only depends on the stop mixing angle $\theta_{\tilde{t}}$. On the other hand, the chargino $\tilde{\chi}^{\pm}_{1}$ and two neutralinos $\tilde{\chi}^{0}_{1,2}$ are nearly degenerate in NSUSY so that the lepton from chargino decay $\tilde{\chi}^{+}_{1} \to W^{+*}\tilde{\chi}^{0}_{1} \to \ell^+ \nu_\ell \tilde{\chi}^0_1$ is too soft to be detected at the LHC. So, we will only discuss the effects of top polarization on kinematics in stop searches in NSUSY at the LHC. It should be noted that in NSUSY, the dark matter candidate -- light higgsino-like neutrolino has a typically low relic density due to the large annihilation rate during the early universe. To be consistent with the observed relic density, several alternatives have been proposed~\cite{Acharya:2010af,Moroi:1999zb,Gelmini:2006pw,Acharya:2007rc,Acharya:2008zi,Choi:2008zq}, such as an axion-higgsino mixing dark matter~\cite{Baer:2011hx,Baer:2011uz}. In this paper, we consider two benchmark points as followings,
\begin{eqnarray}
\textrm{Left-handed (LH) top}:\quad &&\mu=100\textrm{GeV},A_{t}=10,\tan\beta=10,m_{\tilde{Q}_{3L}}=2000 \textrm{GeV}, \nonumber \\ &&m_{\tilde{U}_{3R}}=650\textrm{GeV},m_{\tilde{t}_{1}}=519\textrm{GeV},m_{\tilde{t}_{2}}=2028\textrm{GeV}.\\
\textrm{Right-handed (RH) top:}\quad &&\mu=100\textrm{GeV},A_{t}=10,\tan\beta=10,m_{\tilde{Q}_{3L}}=600\textrm{GeV},\nonumber\\
&&m_{\tilde{U}_{3R}}=2000\textrm{GeV},m_{\tilde{t}_{1}}=518\textrm{GeV},m_{\tilde{t}_{2}}=1988\textrm{GeV}.
\end{eqnarray}
The mass spectrum of sparticles are evaluated by the package~\textsf{SUSYHIT}~\cite{Djouadi:2006bz}. We then generate the parton-level and reconstruction-level events with MadGraph5\_aMC@NLO~\cite{Alwall:2014hca} for two benchmark points at 13 TeV LHC. We implement the parton shower and hadronization by~\textsf{Pythia-8.2}~\cite{pythia} and simulate the detector effects by~\textsf{Delphes}~\cite{delphes}. We cluster the jets by~\textsf{FastJet}~\cite{fastjet} with the anti-$k_t$ algorithm~\cite{anti-kt}.
\par
\begin{figure}[t]
\begin{minipage}{0.48\linewidth}
  \centerline{\includegraphics[scale=0.5]{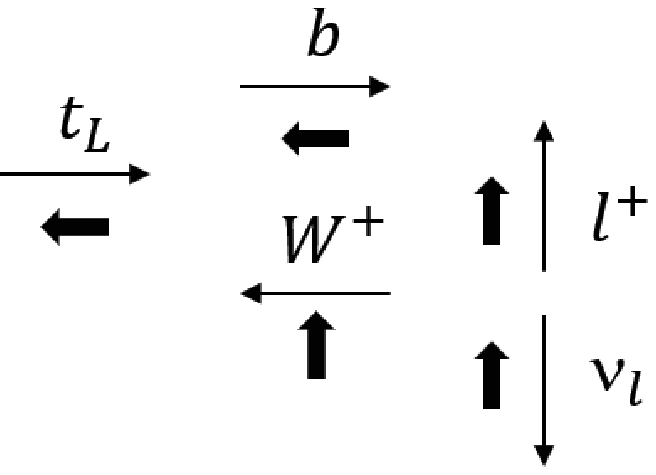}}
  \centerline{(a) LH top with longitudinally polarized $W$.}
\end{minipage}
\hfill
\begin{minipage}{0.48\linewidth}
  \centerline{\includegraphics[scale=0.5]{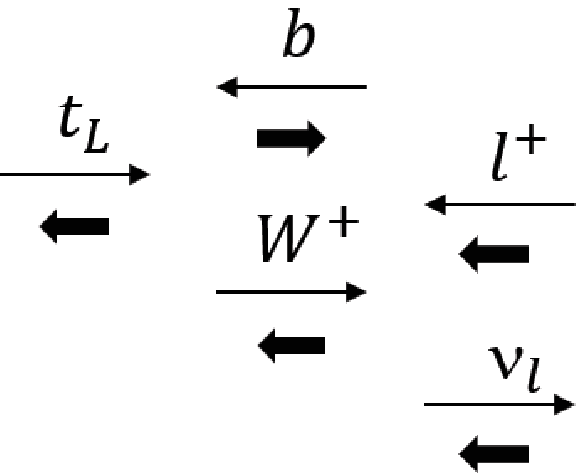}}
  \centerline{(b) LH top with transversely polarized $W$.}
\end{minipage}
\\[10pt]
\begin{minipage}{0.48\linewidth}
  \centerline{\includegraphics[scale=0.5]{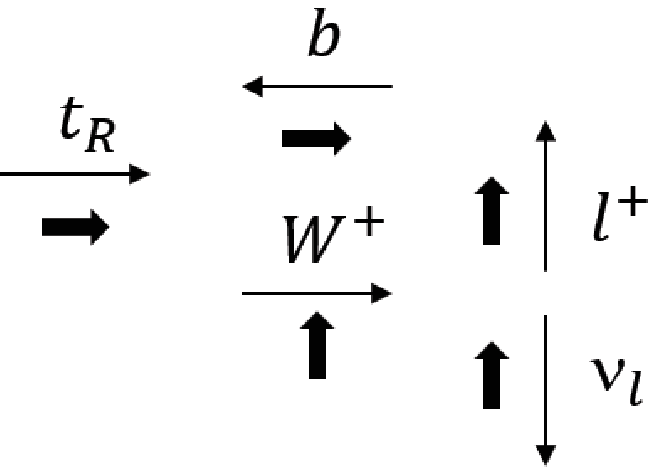}}
  \centerline{(c) RH top with longitudinally polarized $W$.}
\end{minipage}
\hfill
\begin{minipage}{0.48\linewidth}
  \centerline{\includegraphics[scale=0.5]{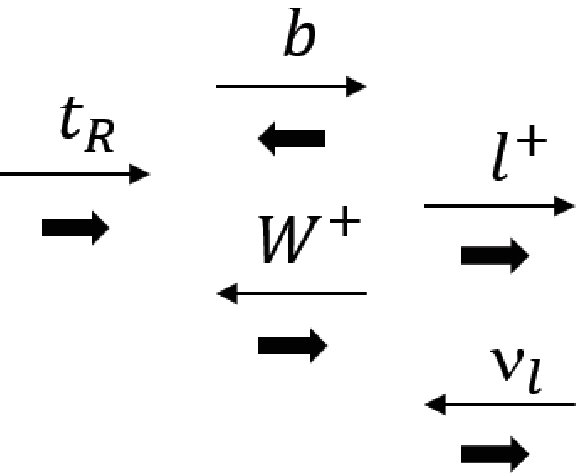}}
  \centerline{(d) RH top with transversely polarized $W$.}
\end{minipage}
\caption{Illustration of helicities for left- and right-handed polarized top decay chains from stop decay. The momenta and spins of particles are denoted by the thin and thick arrow lines, respectively.}
\label{fig:helicity}
\end{figure}
In Fig.~\ref{fig:semi_spin}, we present the kinematical distributions of stop semileptonic channel $pp \to \tilde{t}_1 \tilde{t}^*_1 \to t (\to b j j) \bar{t} (\to \bar{b} \ell^- \bar{\nu}) \tilde{\chi}^0_{1,2} \tilde{\chi}^0_{1,2}$ and $t\bar{t}$ background events at parton- and reconstruction-level at 13 TeV LHC. It can be seen that the transverse momentums of charged lepton ($P_T [\ell]$), $b$-jet ($P_T [b]$), the angular distance between lepton and $b$ jet ($\Delta R [\ell, b]$), the missing transverse energy ($\slashed E_T$) and the azimuthal angle difference ($\Delta \phi [\ell, b]$) for LH and RH top quarks in stop pair events are different from those in $t\bar{t}$ background events. The detector effects do not change the distributions of $P_T [\ell]$ and $P_T [b]$ too much. Besides, the LH top quark shows a different behavior from RH top quark in the distributions of $P_T [\ell]$, $P_T [b]$, $\Delta R [\ell, b]$ and $\Delta\phi[\ell,b]$. For example, leptons in the LH top quark are softer than those in the RH top quark. This can be understood from the analysis of angular momentum conservation, which is shown in Fig.~\ref{fig:helicity}. The successive boosts of the top quark and $W$ boson in the same direction in Fig.~\ref{fig:helicity} (b) with transversely polarized $W$, lead to softer leptons for the LH top quark. While in the case of Fig.~\ref{fig:helicity} (d), boosts of the top quark and $W$ boson in the opposite direction lead to harder leptons in the RH top quark. Similarly, the $b$ quark from LH top quark decay is boosted along its momentum by its parent top quark, as shown in Fig.~\ref{fig:helicity} (a). While for the RH top quark decay, the $b$ quark is boosted against its motion by its parent top quark in Fig.~\ref{fig:helicity}\,(c). Then, we have harder $b$ quarks from the LH top quark than that from RH top quark. \par
From the above discussion, we notice that although the missing transverse energy ($\slashed E_T$) for LH and RH benchmarks are quite close to each other, the transverse momenta $P_{T} [\ell]$, $P_{T} [b]$, the angular distance $\Delta R [\ell,b]$ and azimuthal angle $\Delta\phi [\ell,b]$ can tell the difference between these benchmarks and hard $P_{T} [b]$ or $P_{T} [\ell]$ selection can help distinguish LH top scenario from a RH one. Therefore, the measurements of these variables can be used to identify the nature of interactions between stop and top quark in the natural SUSY.

\section{Conclusions}
In this paper, we studied the polarized top quark and chargino from the stop decays. We calculated the charged lepton angular distribution in the processes $\tilde{t}_1 \to t \tilde{\chi}^0_1 \to (b W^+) \tilde{\chi}^0_1 \to (b \ell^+ \nu_\ell) \tilde{\chi}^0_1$ and $\tilde{t}_1 \to b\tilde{\chi}^+_1 \to b (W^+ \tilde{\chi}^0_1) \to b (\ell^+ \nu \tilde{\chi}^0_1)$ and obtained the corresponding spin-analyzing power for the charged lepton. We found that the mass differences $m_{\tilde{t}_1}-m_{\tilde{\chi}^0_1}$ and $m_{\tilde{\chi}^+_1}-m_{\tilde{\chi}^0_1}$ play an important role in the degrees of polarization of top quark and chargino, respectively. Furthermore, we investigated the top polarization effects on kinematics in searches of stop in natural SUSY at the LHC. We found that several observables, such as the transverse momentums of charged lepton, $b$-jet, the angular distance between lepton and $b$ jet, are sensitive to the polarizations of the top quark and may be useful for probing the nature of stop interactions in natural SUSY after the discovery of top squark.

\section*{Acknowledgement}
This work was supported by the National Natural Science Foundation of China (NNSFC) under grant No. 11705093 and No. 11847208, as well as Jiangsu Specially Appointed Professor Program.

\begin{comment}
%It is then clear that the polarization of the top quark depends totally on the stop mixing angle $\theta_{\tilde{t}}$\,: $\theta_{\tilde{t}}=\pi/2$ corresponds to a left-handed top benchmark and $\theta_{\tilde{t}}=0$
A soft-lepton pre-selection\cite{Aaboud:2017aeu} is applied considering the nearly degeneration of neutralinos and chargino. The signal region requires exactly one signal lepton and no additional baseline leptons. At least two signal jets and one b-tagged jet are also required. $P_{T}> 4\,\textrm{GeV}$ for muons and $P_{T}> 5\,\textrm{GeV}$ for electrons. The signal region relies on $E\!\!\!\!/_{T}$ trigger only and all events should have $E\!\!\!\!/_{T}> 230\,\textrm{GeV}$ to ensure a fully efficient trigger. At the same time, $P_{T}$ of the leading and sub-leading jets should be greater than 25 GeV. These pre-selection criteria are summarized in \tablename~\ref{tab:presel}.
\begin{table}[hbt]
\begin{tabular}{|c|c|}
\hline
Selection  &  Soft-lepton SR\\ \hline
Trigger  &  $E\!\!\!\!/_{T}$ triggers only \\ \hline
No. of leptons  &  1 tight lepton \\ \hline
Lepton $P_{T}$\,[GeV]  &\, > 4 for $\mu$,\, > 5 for e\, \\ \hline
\,No. of (jets, b-tags)  &  ($\geq2,\,\geq1$) \\ \hline
Jet $P_{T}$\,[GeV]  &  > (25,\, 25) \\ \hline
$E\!\!\!\!/_{T}$\,[GeV]  &  > 230 \\
\hline
\end{tabular}
\caption{Pre-selection criteria used for the soft-lepton signal regions.}
\label{tab:presel}
\end{table}
\end{comment}

\end{document}